\documentclass[10pt]{article}

\usepackage{amstext,amsgen,latexsym,amsmath}
\usepackage{amstext,amssymb,amsfonts,latexsym}
\usepackage{theorem}
\usepackage{pifont}
\usepackage{graphics,epsfig}

 \newcommand{\bs}{\bigskip}
 \newcommand{\ms}{\medskip}
 \newcommand{\n}{\noindent}
 \newcommand{\s}{\smallskip}
 \newcommand{\hs}[1]{\hspace*{ #1 mm}}
 \newcommand{\vs}[1]{\vspace*{ #1 mm}}

 \newcommand{\cent}{{|}\!\!\mathrm{c}}

 \newcommand{\nat}{\mathbb{N}}
 
 \newcommand{\integer}{\mathbb{Z}}
 \newcommand{\rational}{\mathbb{Q}}
 
 \newcommand{\complex}{\mathbb{C}}

\newcommand{\algebraic}{\mathbb{A}}

 \newcommand{\bvec}[1]{{\bf #1}}

 \newcommand{\ie}{\textrm{i.e.},\hspace*{2mm}}
 \newcommand{\eg}{\textrm{e.g.},\hspace*{2mm}}
 \newcommand{\etal}{\textrm{et al.}\hspace*{2mm}}
 \newcommand{\etalc}{\textrm{et al.}}

 \newcommand{\AAA}{{\cal A}}
 
 \newcommand{\CC}{{\cal C}}

 \newcommand{\MM}{{\cal M}}

 \newcommand{\PP}{{\cal P}}
 
 \newcommand{\RR}{{\cal R}}
 
 \newcommand{\VV}{{\cal V}}

 \newcommand{\p}{\mathrm{{P}}}
 \newcommand{\np}{\mathrm{{NP}}}

 \newcommand{\am}{\mathrm{{AM}}}

 \newcommand{\pspace}{\mathrm{{PSPACE}}}

 \newcommand{\qip}{\mathrm{QIP}}
 \newcommand{\ip}{\mathrm{IP}}

 \def\bbox{\vrule height6pt width6pt depth1pt}

\theoremstyle{plain}
\theoremheaderfont{\bfseries}
 \newtheorem{theorem}{Theorem}[section]
 \newtheorem{lemma}[theorem]{Lemma}
 \newtheorem{proposition}[theorem]{Proposition}

 {\theorembodyfont{\rmfamily}
\newtheorem{definition}[theorem]{Definition}}
 {\theorembodyfont{\rmfamily} }
 {\theorembodyfont{\rmfamily} }

 \newenvironment{proof}{\par \noindent
            {\bf Proof. \hs{2}}}{\hfill$\Box$ \vspace*{3mm}}

 \newenvironment{proofof}[1]{\vspace*{5mm} \par \noindent
         {\bf Proof of #1.\hs{2}}}{\hfill$\Box$ \vspace*{3mm}}

 \newcommand{\ceilings}[1]{\lceil #1 \rceil}
 
 \newcommand{\pair}[1]{\langle #1 \rangle}

 \newcommand{\qubit}[1]{| #1 \rangle}

 \newcommand{\ketbra}[2]{| #1 \rangle\!\langle #2 |}

\newif\ifnotesw\noteswtrue
   {\ifnotesw\marginpar[\hfill\(\top\)]{\(\top\)}\fi}%
      {\ifnotesw\marginpar[\hfill\(\bot\)]{\(\bot\)}\fi}
      
\newcommand{\mnote}[1]%
   {\ifnotesw\marginpar%
	  [{\scriptsize\begin{minipage}[t]{\marginparwidth}
	  \raggedleft#1%
		  \end{minipage}}]%
	  {\scriptsize\begin{minipage}[t]{\marginparwidth}
	  \raggedright#1%
		  \end{minipage}}%
    \fi}

\newcommand{\ignore}[1]{}

\begin{document} 
\pagestyle{plain} 
\begin{center}
{\Large {\bf An Application of Quantum Finite Automata \s\\
to Interactive Proof Systems}}
\footnote{An extended abstract appeared in the Proceedings 
of the 9th International Conference on Implementation and 
Application of Automata, 
Lecture Notes in Computer Science, Springer-Verlag, Kingston, 
Canada, July 22--24, 2004. 
This work was in part supported by the Natural Sciences and 
Engineering Research Council of Canada.} \bs\\
\begin{tabular}{c@{\hspace{20mm}}c} 
{\sc Harumichi Nishimura} & {\sc Tomoyuki Yamakami} \\
\end{tabular}\bs\\ 
{Computer Science Program, Trent University \\
Peterborough, Ontario, Canada K9J 7B8}  
\end{center} 

\bs
\n {\bf Abstract}: 
Quantum finite automata have been studied intensively since 
their introduction in late 1990s as a natural model of a 
quantum computer with finite-dimensional quantum memory space. 
This paper seeks their direct application  
to interactive proof systems in which a mighty quantum prover 
communicates with a quantum-automaton verifier 
through a common communication cell. Our quantum interactive 
proof systems are juxtaposed to Dwork-Stockmeyer's classical 
interactive proof systems whose verifiers are two-way 
probabilistic automata. We demonstrate strengths and weaknesses 
of our systems and further study how various restrictions 
on the behaviors of quantum-automaton verifiers affect the power 
of quantum interactive proof systems.
\ms

\n{\sf Keywords:} quantum finite automaton, quantum interactive 
proof system, quantum measurement, quantum circuit

\section{Development of Quantum Finite Automata}\label{sec:QFA} 

A quantum computer---quantum-mechanical computing device---has 
drawn wide attention 
as a future computing paradigm since the pioneering work of 
Feynman \cite{Fey82}, Deutsch \cite{Deu85}, and Benioff \cite{Ben80} 
in the 1980s. Over the decades, such a device has been 
mathematically modeled in numerous ways to deliver a coherent 
theory of quantum computation. Of all computational models,  
Moore and Crutchfield \cite{MC00} as well as Kondacs and 
Watrous \cite{KW97} 
proposed a (one-head) {\em quantum finite automaton} ({\em qfa}, 
in short) as a simple but natural 
model of a quantum computer that is equipped with finite-dimensional 
quantum memory space\footnote{The tape head of a quantum finite 
automaton may exist in a superposition.}. Parallel to classical 
automata theory, the theory of quantum finite automata has been 
well established to study the nature of quantum computation.
Performing a series of unitary operations as its tape head scans 
input symbols, a qfa may eventually enter 
accepting or rejecting inner states to halt. 
Any entry of such a unitary operation is a complex number, 
called a {\em (transition) amplitude}. 
A quantum computation is seen as an evolution of a quantum 
superposition of the machine's configurations, where 
a configuration is a pair of an inner state and a head position 
of the machine. 
As quantum physics dictates, a quantum evolution is reversible 
in nature. 
A special operation called a {\em (quantum) measurement} is 
performed to {\lq\lq}observe{\rq\rq} 
whether the qfa enters an accepting inner state, a rejecting inner 
state, or a non-halting inner state. 
Of all the variations of qfa's discussed in the past literature, 
we shall focus our study only 
on the early models of Moore and Crutchfield and of Kondacs and 
Watrous for our application to interactive proof systems. 

In 1997, Kondacs and Watrous \cite{KW97} introduced two types of 
qfa's: a {\em 1-way quantum finite automaton} ({\em 1qfa}, in short) 
whose head always moves rightward and a {\em 2-way quantum finite 
automaton} ({\em 2qfa}, in short) whose head moves in all 
directions. Both qfa's perform a so-called projection 
measurement (or von Neumann measurement) after every move of them. 
Because of a finite memory constraint, no 1qfa recognizes even 
the regular language $Zero=\{x0\mid x\in\{0,1\}^*\}$ with small 
error probability \cite{KW97}. 
In the model of Moore and Crutchfield, on the contrary, a 1qfa 
performs a measurement only once after the tape head scans the 
right endmarker. 
Their model is often referred to as a {\em measure-once 1-way 
quantum finite automaton} ({\em mo-1qfa}, in short). The qfa 
model of Kondacs and Watrous is by contrast called a {\em measure-many 
1-way quantum finite automaton}. 
As Brodsky and Pippenger \cite{BP02} showed, mo-1qfa's are so 
restrictive that they are fundamentally equivalent in power 
to {\lq\lq}permutation{\rq\rq} 
automata, which recognize exactly group languages.  
Unlike the 1qfa's, 2qfa's can simulate deterministic finite 
automata with probability $1$. Moreover, 
Kondacs and Watrous \cite{KW97} constructed a 2qfa that recognizes 
with small error probability the non-regular language 
$Upal=\{0^n1^n\mid n\geq0\}$ (unique palindromes) in worst-case 
linear time by exploiting its quantum superposition.  
The power of a qfa may vary in general depending on the types 
of restrictions imposed on its behaviors: 
for instance, head move, measurement, quantum state, and so forth.

We are particularly interested in a qfa whose error probability 
is bounded above 
by a certain constant $\epsilon \in[0,1/2)$ independent of input 
lengths. 
Such a qfa is conventionally called {\em bounded error}. 
We use the notation $\mathrm{1QFA}$ ($\mathrm{2QFA}$, resp.) 
to denote the class of all languages recognized by bounded-error 
1qfa's (2qfa's, resp.) with arbitrary complex amplitudes. 
Similarly, let 
$\mathrm{MO\mbox{-}1QFA}$ be the class of all languages recognized 
by bounded-error mo-1qfa's. When the running time of a qfa is 
an issue, we use the notation $\mathrm{2QFA}(poly\mbox{-}time)$ to 
denote the collection of all languages recognized by expected 
polynomial-time 2qfa's with bounded error, where an {\em expected 
polynomial-time 2qfa} is a 2qfa whose average running time on each 
input of length $n$ is bounded above by a fixed polynomial in $n$. 
When all amplitudes are drawn from a designated amplitude set $K$, 
we emphatically write $\mathrm{2QFA}_K$ and 
$\mathrm{2QFA}_K(poly\mbox{-}time)$. For comparison, we write 
$\mathrm{REG}$ for the class of all regular languages. 
Our current state of knowledge is summarized as follows: 
$\mathrm{1QFA}\subsetneqq \mathrm{REG} \subsetneqq 
\mathrm{2QFA}(poly\mbox{-}time) \subseteq \mathrm{2QFA}$. 
How powerful is $\mathrm{2QFA}$? 
It directly follows from \cite{Wat99} that any 2qfa with 
$\algebraic$-amplitudes\footnote{The set $\algebraic$ consists 
of all algebraic complex numbers.} 
can be simulated by a probabilistic Turing machine (PTM, in short) 
using space $O(\log{n})$ with unbounded error. 
Since any unbounded-error $s(n)$-space PTM can be simulated 
deterministically in time $2^{O(s(n))}$ \cite{BCP83}, 
we conclude that $\mathrm{2QFA}_{\algebraic}\subseteq \p$. 
For an overview of qfa's, see the textbook, \eg \cite{Gru99}.

In this paper, we seek a direct application of qfa's to an 
interactive proof system, which can be viewed as a two-player 
game between the 
players called a prover and a verifier. In our basic model, 
a qfa plays a role of a verifier and a prover can apply any 
operation that quantum physics allows. Such a system is generally 
called a weak-verifier quantum interactive proof system. We further 
place various restrictions on our basic model and study how such 
restrictions affect its computational power. In the following 
section, we take a quick tour of the notion of 
interactive proof systems as an introduction to our formalism 
of quantum interactive proof systems with qfa verifiers.

\section{Basics of Interactive Proof Systems}\label{sec:basics}

In mid 1980s, Goldwasser, Micali, and Rackoff \cite{GMR89} and 
independently Babai \cite{Bab85} 
introduced the notion of a so-called (single-prover) {\em interactive 
proof system} 
({\em IP system}, in short), which can be viewed as a two-player 
game in which a player $P$, called a {\em prover}, 
who has unlimited computational power tries to convince or fool 
the other player $V$, 
called a {\em verifier}, who runs a randomized algorithm. These 
two players can access a given input and share a common communication 
bulletin board on which they can communicate with each other by 
posting their messages in turn. The goal of the verifier is to decide 
whether the input is in a given language $L$ with designated accuracy. 
We say that {\em $L$ has an IP system} $(P,V)$ (or {\em an IP system 
$(P,V)$ recognizes} $L$) if there exists an error bound 
$\epsilon\in[0,1/2)$ such that the following two conditions hold: 
(1) if the input $x$ belongs to $L$, then the {\lq\lq}honest{\rq\rq} 
prover $P$ convinces the verifier $V$ to accept $x$ with probability 
$\geq 1-\epsilon$ and (2) if the input $x$ is not in $L$, then the 
verifier $V$ rejects $x$ with probability $\geq 1-\epsilon$ although 
it plays against any {\lq\lq}dishonest{\rq\rq} prover. Because 
of their close connection to cryptography, program checking, and list 
decoding, the IP systems have become one of the major research 
topics in computational complexity theory.

When a verifier is a polynomial-time PTM, Shamir \cite{Sha} proved 
that the corresponding IP 
systems exactly characterize the complexity class $\pspace$ based on 
the work of Lund, Fortnow, Karloff, 
and Nisan \cite{LFKN92} and on the result of Papadimitriou \cite{Pap85}. 
This demonstrates the power of interactions between mighty provers and 
polynomial-time PTM verifiers.

The major difference between the models of Goldwasser \etalc~\cite{GMR89} 
and of Babai \cite{Bab85} is the amount of the verifier's private 
information that is revealed to a prover. Goldwasser \etal considered 
the IP systems whose verifiers can hide his probabilistic moves from 
provers to prevent any malicious attack of the provers. Babai 
considered by contrast the IP systems in which verifiers' moves are 
completely revealed to provers. Although he named his IP system an 
{\em Arthur-Merlin game}, it is also known as an IP system with 
{\lq\lq}public coins.{\rq\rq} 
Despite the difference of the models, Goldwasser and Sipser \cite{GS86} later
proved that the classes of all languages recognized by both IP systems 
with polynomial-time PTM verifiers coincide. 

In early 1990s, Dwork and Stockmeyer \cite{DS92} focused their research 
on IP systems with weak verifiers, particularly, bounded-error {\em 2-way 
probabilistic finite automaton} ({\em 2pfa}, in short) verifiers that 
may {\lq\lq}privately{\rq\rq} flip fair coins. Their research inspires 
us to apply quantum finite automata to interactive proof systems. 
For later use, let $\ip(2pfa)$ be the class of all languages recognized 
by IP systems with 2pfa verifiers 
and let $\ip(2pfa,poly\mbox{-}time)$ be the subclass of $\ip(2pfa)$ 
where the verifiers run in expected polynomial time. 
When the verifiers flip only {\lq\lq}public coins,{\rq\rq} we write 
$\am(2pfa)$ and 
$\am(2pfa,poly\mbox{-}time)$ instead. Dwork and Stockmeyer showed without 
any unproven assumption that the IP systems with 2pfa verifiers are more 
powerful than 2pfa's alone (which are viewed as IP systems without any 
prover). Moreover, they showed that the non-regular language 
$Pal=\{x\in\{0,1\}^*\mid x=x^{R}\}$ (palindromes), where $x^{R}$ is 
$x$ in the reverse order, separates $\ip(2pfa,poly\mbox{-}time)$ from 
$\am(2pfa)$ 
and the language $Center=\{x1y\mid x,y\in\{0,1\}^*,|x|=|y|\}$ separates 
$\am(2pfa)$ from $\am(2pfa,poly\mbox{-}time)$. 
The IP systems of Dwork and Stockmeyer can be seen as a special case 
of a much broader concept of 
space-bounded IP systems. For their overview, the reader may refer 
to \cite{Con93}.

Recently, a quantum analogue of an IP system was introduced by 
Watrous \cite{Wat03} under the term (single-prover) {\em quantum 
interactive proof system} ({\em QIP system}, in short). The QIP systems 
with uniform polynomial-size quantum-circuit verifiers exhibit 
significant computational power of recognizing every language in 
$\mathrm{PSPACE}$ by exchanging only three messages between a prover 
and a verifier \cite{KW00,Wat03}. The study of QIP systems, including 
their variants (such as multi-prover model \cite{CHTW04,KM03} and 
zero-knowledge model \cite{Kob03,Wat02}), 
has become a major topic in quantum complexity theory.  In particular, 
quantum analogues of Babai's Merlin-Arthur games, called {\em quantum 
Merlin-Arthur games},  
have drawn significant attention (e.g., \cite{AN02,AR03,KMY03,Wat00,Yao03}). 

Motivated by the work of Dwork and Stockmeyer \cite{DS92}, this paper 
introduces a QIP system whose verifier is especially a qfa. In the 
subsequent sections, we give the formal definition of our basic QIP 
systems and explore their properties and relationships to the classical 
IP systems of Dwork and Stockmeyer.

\section{Application of QFAs to QIP Systems}\label{application}

Following the success of IP systems with 2pfa verifiers, we wish to 
apply qfa's to QIP 
systems. A purpose of our study is to examine the power of 
{\lq\lq}interaction{\rq\rq} 
when a weak verifier, represented by a qfa, meets with a mighty prover. 
The main goal of our study is (i) to investigate the roles of the interactions 
between a prover and a weak verifier, (ii) to understand the influence of 
various restrictions and extensions of QIP systems, and (iii) to study 
the QIP systems under a broader but general framework. In addition, 
when the power of 
verifiers is limited, we may possibly prove without any unproven assumption the 
separations and collapses of certain complexity classes defined by QIP systems 
with such weak verifiers.

Throughout this paper, let $\rational$ and $\complex$ respectively denote 
the sets of all rational numbers and of all complex numbers. 
Let $\nat$ be the set of all natural numbers (\ie nonnegative integers) 
and set $\nat^{+}= \nat-\{0\}$. For any two integers $m$ and $n$ with 
$m<n$, the notation $[m,n]_{\integer}$ denotes the set 
$\{m,m+1,m+2,\ldots,n\}$ and $\integer_{n}$ in particular 
denotes the set $[0,n-1]_{\integer}$. All {\em logarithms} are to base 2 
and all {\em polynomials} have integer coefficients. By $\tilde{\complex}$, 
we denote 
the set of all polynomial-time approximable complex numbers, where a complex 
number is called {\em polynomial-time approximable} if its real part and 
imaginary part 
are both deterministically approximated to within 
$2^{-n}$ in polynomial time. Our input alphabet $\Sigma$ 
is an arbitrary finite set, not necessarily limited to $\{0,1\}$. 
Following the convention, we write $\Sigma^{n}=\{x\in\Sigma^*\mid |x|=n\}$ 
and $\Sigma^{\leq n}=\{x\in\Sigma^*\mid |x|\leq n\}$, where $|x|$ 
denotes the length of $x$. Opposed to the notation $\Sigma^*$,  
$\Sigma^{\infty}$ stands for the collection of all infinite sequences, 
each 
of which consists of symbols from $\Sigma$. For any symbol $a$ in 
$\Sigma$, $a^{\infty}$ denotes an element of $\Sigma^{\infty}$, 
which is the infinite sequence made only of $a$.
We assume the reader's familiarity with classical automata theory 
and the basic concepts of quantum 
computation (see, \eg \cite{Gru99,HMU01,NC00}). 

\subsection{Basic Definition}\label{sec:basic-def}

We first give a {\lq\lq}basic{\rq\rq} definition of a QIP system 
whose verifier is a qfa. 
Our basic definition is a natural concoction of the IP model of Dwork 
and Stockmeyer \cite{DS92} and the qfa model of Kondacs and Watrous 
\cite{KW97}. 
In the subsequent section, we discuss a major difference between our 
QIP systems and the circuit-based QIP systems 
of Watrous \cite{Wat03}. Our definition seemingly demands much stricter 
conditions than that of Dwork and Stockmeyer; however, our basic 
model serves a mold to build various QIP systems with qfa verifiers. 
In later sections, we shall restrict the behaviors of a 
verifier as well as a prover to obtain several variants of our basic QIP 
systems since these restricted models have never been addressed in 
the literature. 

Hereafter, the notation 
$(P,V)$ is used to denote the QIP system with the prover $P$ and the 
verifier $V$. In such a QIP system $(P,V)$, the 2qfa verifier $V$ 
is particularly
specified by a finite set $Q$ of verifier's inner states, a finite input 
alphabet $\Sigma$, a finite communication alphabet $\Gamma$, and a 
verifier's transition function $\delta$. The set $Q$ is the union of 
three mutually disjoint subsets $Q_{non}$, $Q_{acc}$, and $Q_{rej}$, 
where any states in $Q_{non}$, $Q_{acc}$, and $Q_{rej}$ are respectively 
called a {\em non-halting inner state}, an {\em accepting inner state}, 
and a 
{\em rejecting inner state}. Accepting inner states and rejecting 
inner states are simply 
called {\em halting inner states}. In particular, $Q_{non}$ has the 
so-called 
{\em initial inner state} $q_0$. The input tape is indexed by natural 
numbers (the first cell is indexed $0$). The two designated symbols 
$\cent$ and $\$$ not in $\Sigma$, called respectively the 
{\em left endmarker}\footnote{For certain variants of qfa's, the 
left endmarker is 
redundant. See, \eg \cite{ABG+04}.} and the {\em right endmarker}, 
mark the left end and the right end of the input. For convenience, set
 $\check{\Sigma}=\Sigma\cup\{\cent,\$\}$. Assume also that $\Gamma$ 
contains the blank symbol $\#$. At the beginning of the computation, 
an input string $x$ over $\Sigma$ of length $n$ is written orderly from 
the first cell to the $n$th cell of the input tape. The tape head initially 
scans the left endmarker. The communication cell holds only a symbol 
in $\Gamma$ and initially the blank symbol $\#$ is written in the 
cell. Similar to the original definition of \cite{KW97}, our input tape 
is {\em circular}; that is, whenever the verifier's head scanning $\cent$ 
($\$$, resp.) on the input tape moves to the left (right, resp.), the 
head reaches to the right end (resp. left end) of the input tape. 

\begin{figure}[ht]
\bs
\begin{center}
\centerline{\psfig{figure=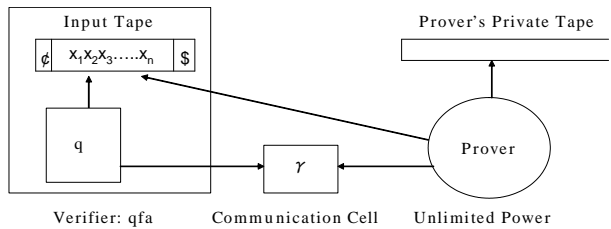,height=3.0cm}}
\caption{A schematic of a QIP system with a qfa verifier}\label{fig:scheme}
\end{center}
\end{figure}

A {\em (global) configuration of $(P,V)$} is a description of the QIP 
system $(P,V)$ at a certain moment, comprising visible configurations 
of the two players. Each player can see only his portion of a global 
configuration. 
A {\em visible configuration} of the verifier $V$ on an input of 
length $n$ is 
represented by a triplet 
$(q,k,\gamma)\in Q\times\integer_{n+2}\times\Gamma$, 
which indicates that the verifier is in state $q$, the content of the 
communication cell is $\gamma$, and the verifier's head position is 
$k$ on the input tape. Let $\VV_n$ and $\MM$ be respectively the Hilbert 
spaces spanned by the computational bases 
$\{\qubit{q,k}\mid (q,k)\in Q\times\integer_{n+2}\}$ and 
$\{\qubit{\gamma}\mid \gamma\in\Gamma\}$. The Hilbert space 
$\VV_n\otimes\MM$ is called the {\em verifier's visible configuration 
space} on inputs of length $n$. The {\em verifier's transition 
function} $\delta$ is a map from $Q\times\check{\Sigma}\times\Gamma\times 
Q\times\Gamma\times\{0,\pm 1\}$ to $\complex$ and is interpreted as 
follows. For any $q,q'\in Q$, $\sigma\in\check{\Sigma}$, 
$\gamma,\gamma'\in\Gamma$, 
and $d\in\{0,\pm 1\}$, the complex number 
$\delta(q,\sigma,\gamma,q',\gamma',d)$ 
specifies the transition amplitude with which the verifier $V$ 
scanning symbol 
$\sigma$ on the input tape and symbol $\gamma$ on the communication 
cell in state $q$ changes $q$ to $q'$, replaces $\gamma$ with $\gamma'$, 
and moves the machine's head on the input tape in direction $d$.  

For any input $x$ of length $n$, $\delta$ induces the linear 
operator $U_\delta^x$ on $\VV_n\otimes\MM$ defined by 
$U_\delta^x\qubit{q,k,\gamma}= \sum_{q',\gamma',d} 
\delta(q,x_{(k)},\gamma,q',\gamma',d)\qubit{q',k',\gamma'}$, where 
$x_{(k)}$ is the $k$th symbol in $x$ and $k'=k+d\ (\mbox{mod }n+2)$. 
The verifier is called {\em well-formed} if $U_\delta^x$ is 
unitary on $\VV_n\otimes\MM$ for every string $x\in\Sigma^*$. Since 
we are interested only in well-formed 
verifiers, we henceforth assume that all verifiers are well-formed. 
For every input $x$ of length $n$, the 2qfa verifier $V$ starts with 
the initial superposition $\qubit{q_0,0,\#}$. A single step of the verifier 
on input $x$ consists of the following process. First, $V$ applies 
his operation $U_\delta^x$ to an existing superposition $\qubit{\phi}$ 
and then $U_\delta^x\qubit{\phi}$ becomes the new superposition 
$\qubit{\phi'}$. Let $W_{acc}=\mathrm{span}\{\qubit{q,k,\gamma}\mid 
(q,k,\gamma)\in Q_{acc}\times\integer_{n+2}\times\Gamma\}$, 
$W_{rej}=\mathrm{span}\{\qubit{q,k,\gamma}\mid (q,k,\gamma)\in Q_{rej}\times\integer_{n+2}\times\Gamma\}$, 
and $W_{non}=\mathrm{span}\{\qubit{q,k,\gamma}\mid (q,k,\gamma)\in Q_{non}\times\integer_{n+2}\times\Gamma\}$. Moreover, let $k_{acc}$, 
$k_{rej}$, and $k_{non}$ be respectively the positive
numbers representing {\lq\lq}accept,{\rq\rq} 
{\lq\lq}reject,{\rq\rq} and {\lq\lq}non halt.{\rq\rq} 
The new superposition $\qubit{\phi'}$ is then measured by the observable 
$k_{acc}E_{acc}+ k_{rej}E_{rej}+ k_{non}E_{non}$, where $E_{acc}$, 
$E_{rej}$, and $E_{non}$ are respectively the projection operators 
on $W_{acc}$, $W_{rej}$, and $W_{non}$.  
Provided that $\qubit{\phi'}$ is expressed as 
$\qubit{\psi_1}+\qubit{\psi_2}+\qubit{\psi_3}$ for certain three vectors 
$\qubit{\psi_1}\in W_{acc}$, $\qubit{\psi_2}\in W_{rej}$, and 
$\qubit{\psi_3}\in W_{non}$, we say that, at this step, {\em $V$ 
accepts $x$ with probability} $\|\qubit{\psi_1}\|^2$ and {\em  
rejects $x$ with probability} $\|\qubit{\psi_2}\|^2$. Only the 
non-halting superposition $\qubit{\psi_3}$  
continues to the next step and $V$ is said to {\em continue (to 
the next step) with probability} $\|\qubit{\psi_3}\|^2$. The 
probability that $x$ is accepted (rejected, resp.) within the first 
$t$ steps is thus the sum, over all $i\in[1,t]_{\integer}$, of 
the probabilities with which $V$ accepts (rejects, resp.) $x$ at 
the $i$th step. In particular, when 
the verifier is a 1qfa, the verifier's transition function $\delta$ 
must satisfy the following two additional conditions: (i) for 
every $q,q'\in Q$, 
$\sigma\in\check{\Sigma}$, and $\gamma,\gamma'\in\Gamma$, 
$\delta(q,\sigma,\gamma,q',\gamma',d) =0$ if $d\neq 1$ (i.e., the 
head always moves to the right) and (ii) the verifier must enter halting 
states until the verifier's head moves off the right endmarker $\$$ 
(the head may halt at $\cent$ since the input tape is circular). 
This second condition makes all computation paths terminate. 
Therefore, on input $x$, a 1qfa verifier halts in at most $|x|+2$ steps.

In contrast to the verifier, the prover $P$ has an infinite private 
tape and accesses an input $x$ and a communication cell. Let $\Delta$ 
be a finite set of the prover's private tape alphabet, which 
includes the blank 
symbol $\#$. The  
prover is assumed to alter only a {\lq\lq}finite{\rq\rq} initial 
segment of his private tape at every step. Let $\PP$ be the 
Hilbert space spanned by $\{\qubit{y}\mid y\in \Delta^{\infty}_{fin}\}$, 
where $\Delta^{\infty}_{fin}$ is the set of all finite series 
of tape symbols containing only a finite number of non-blank 
symbols; namely, $\Delta^*\times\{\#\}^{\infty}$. The {\em prover's 
visible configuration space} is the Hilbert space 
$\MM\otimes\PP$. 
Formally, the prover $P$ on input $x$ is specified by a series 
$\{U_{P,i}^x\}_{i\in\nat^+}$ of unitary 
operators, each of which acts on the  prover's visible
configuration space, such that $U_{P,i}^x$ is of the form 
$S_{P,i}^x\otimes I$, where $\dim(S_{P,i}^x)$ is finite and $I$ is the 
identity operator. Such a series of operators is particularly called the 
{\em prover's strategy} on the input $x$. To refer to the 
strategy on $x$, we often use the notation $P_x$. For any function 
$k$ from $\nat^2$ to $\nat$, 
we call the prover {\em $k(n,i)$-space bounded} if the prover 
uses at 
most the first $k(n,i)$ cells of his private tape; that is, at 
the $i$th step, $S_{P,i}^x$ 
is applied only to the first $k(n,i)$ cells of the prover's private 
tape in addition to the 
communication cell. We often consider the case where the value 
$k(n,i)$ is independent of $i$. If the prover has a string $y$ 
in his private tape 
and scans symbol $\gamma$ in the communication cell, then he applies 
$U_{P,i}^x$ to the quantum state $\qubit{\gamma}\qubit{y}$ at the 
$i$th step. If $U_{P,i}^x\qubit{\gamma}\qubit{y} 
=\sum_{\gamma',y'}\alpha_{\gamma',y'}^{i}\qubit{\gamma'}\qubit{y'}$, 
then the prover changes $y$ into $y'$ and replaces $\gamma$ by 
$\gamma'$ 
with amplitude $\alpha_{\gamma',y'}^{i}$. 

Formally, a global configuration consists of the four items: 
$V$'s inner state, $V$'s head position, 
the content of a communication cell, and the content of $P$'s 
private tape. We express a superposition of such configurations 
of $(P,V)$ on input $x$ as
a vector in the Hilbert space $\VV_{|x|}\otimes\MM\otimes\PP$, 
which is called the {\em (global) configuration space} of $(P,V)$ 
on input $x$.  The computation of $(P,V)$ on input $x$ constitutes 
a series of superpositions of configurations resulting 
by an alternate application of unitary operations of the verifier 
and the prover as well as the verifier's measurement. 
The computation on input $x$ starts with the global initial 
configuration $\qubit{q_0,0}\qubit{\#}\qubit{\#^{\infty}}$, in 
which the verifier is in his initial configuration and the prover's 
private tape 
consists only of blank symbols. The two players apply their unitary 
operations $U_\delta^x$ and $P_x=\{U_{P,i}^x\}_{i\in\nat^+}$ in 
turn starting 
with the verifier's move. Through the communication cell, the 
two players exchange communication symbols, 
which cause the two players entangled. A measurement is made 
after every move of 
the verifier to determine whether $V$ is in a halting inner state. 
Each 
computation path therefore ends when $V$ enters a certain 
halting inner state 
along this computation path. 
For convenience, we use the same notation $(P,V)$ to mean a QIP 
system 
and also a protocol taken by the prover $P$ and the verifier 
$V$.  Furthermore, we define 
the overall probability that $(P,V)$ {\em accepts} 
({\em rejects}, resp.) the input $x$ as the 
limit, as $t\rightarrow\infty$, of the probability that $V$  
accepts (rejects, resp.) $x$ in at most $t$ steps. We use the 
notation $p_{acc}(x,P,V)$ ($p_{rej}(x,P,V)$, resp.) to denote 
the overall acceptance (rejection, resp.) probability of $x$ 
by $(P,V)$. We say 
that $V$ {\em always halts with probability $1$} if, for every 
input 
$x$ and every prover $P^{*}$, $(P^{*},V)$ reaches halting inner 
states with 
probability $1$. In general, $V$ may not always halt with probability 
$1$. When we discuss the entire {\em running time} of the QIP 
system, we count the 
number of all steps taken by the verifier as well as the prover.

Let $a,b$ be any two real numbers in the unit 
interval $[0,1]$ and let $L$ be any language. We say that {\em $L$ 
has an $(a,b)$-QIP system} $(P,V)$ (or {\em a $(a,b)$-QIP system 
$(P,V)$ recognizes} $L$) 
if $(P,V)$ is a QIP system and the following two conditions hold 
for $(P,V)$:
\begin{enumerate}
\item {\sf (completeness)} for any $x\in L$, $(P,V)$ accepts $x$ 
with probability at least $a$, and 
\item {\sf (soundness)} for any $x\not\in L$ and any prover $P^*$, 
$(P^*,V)$ rejects\footnote{Generally, the QIP system may increase 
its power if we 
instead require $(P^*,V)$ to {\em accept} $x$ with probability 
$\leq 1-b$ for any prover $P^*$. Such a modification defines a 
{\em weak} QIP system. See, \eg \cite{DS92} for the classical 
case.} $x$ with probability at least $b$. 
\end{enumerate}

Note that a $(a,a)$-QIP system has the error probability at most 
$1-a$. This paper discusses only the QIP systems whose error 
probabilities are bounded above by certain constants lying in 
the interval $[0,1/2)$.

Adapting the notational convention of  Condon \cite{Con93}, we write 
$\qip_{a,b}(\pair{\RR})$, where $\pair{\RR}$ is a set of restrictions, 
to denote the collection of all languages recognized by certain 
$(a,b)$-QIP systems with the restrictions specified by $\pair{\RR}$. 
Let $\qip(\pair{\RR})$ be 
$\bigcup_{\epsilon>0}\qip_{1/2+\epsilon,1/2+\epsilon}(\pair{\RR})$. If 
in addition the verifier's amplitudes are restricted to an amplitude 
set 
$K$ (but there is no restriction for the prover), then we rather 
write $\qip_{K}(\pair{\RR})$. Notice that 
$\qip(\pair{\RR})=\qip_{\complex}(\pair{\RR})$. Mostly, we focus 
our attention on the following three basic restrictions 
$\pair{\RR}$: $\pair{1qfa}$ 
({\lq\lq}measure-many{\rq\rq} 1qfa verifiers), $\pair{2qfa}$ 
({\lq\lq}measure-many{\rq\rq} 2qfa verifiers), and 
$\pair{poly\mbox{-}time}$ (expected polynomial running time). For 
instance, $\qip(2qfa,poly\mbox{-}time)$ denotes the language class 
defined by QIP systems with expected polynomial-time 2qfa verifiers. 
Other types of restrictions will be discussed in later sections.

\subsection{Comparison with Circuit Based QIP Systems}
\label{sec:circuit}

We briefly discuss the major difference between our automaton-based 
QIP systems and circuit-based QIP systems in which a prover and a 
verifier are both viewed as two finite series of quantum circuits 
intertwined each 
other in turn, sharing only message qubits. Here, assumed is 
the reader's 
familiarity with Watrous's circuit-based QIP model \cite{Wat03}. 

In the circuit-based model of Watrous, the measurement of the output 
qubit is performed only once at the end of the computation 
since any measurement during the computation can be postponed to 
the end (see, \eg \cite{NC00}). This is possible because the 
verifier uses his own 
private qubits and his running time is bounded. However, since our 
2qfa verifier has no private tape and may not halt within a finite 
number 
of steps, the simulation of such a verifier on a quantum circuit 
requires a measurement of a certain number of qubits (as a halting flag) 
after each move of the verifier. 

A verifier in the circuit-based model is allowed to carry out a large 
number of basic unitary operations in its single interaction round 
whereas a qfa verifier in our basic 
model is constantly under attack of a malicious prover after every 
move of 
the verifier. This comes from the belief that no malicious prover 
truthfully keeps the communication cell unchanged 
while awaiting for the verifier's next query. Therefore, such a 
malicious prover may exercise more influence on the verifier in our QIP 
model than in the circuit-based model. Later in Section 
\ref{sec:interaction}, 
nevertheless, we shall introduce a variant of our basic QIP systems, 
in which we allow a 
verifier to make a series of transitions without communicating with 
a prover. This makes it possible for us to discuss the number of 
communications between a prover and a verifier necessary for the 
recognition of a given language.

\section{One-Way QFA Verifiers against Mighty Provers}

Following the definition of a qfa-verifier QIP systems, we shall 
demonstrate how well a qfa verifier plays against a powerful 
prover. We begin with our investigation on the power of QIP systems 
whose verifiers are particularly limited to 1qfa's. 

Earlier, Kondacs and Watrous \cite{KW97} demonstrated a weakness of 
1qfa's; namely, no 1qfa recognizes the regular language $Zero$ 
and therefore, $\mathrm{1QFA}$ cannot contain $\mathrm{REG}$. In the 
following theorem, 
we show that the interaction between a prover and a 1qfa verifier 
complements such deficiency of 1qfa's and 
truly enhances the power of recognizing languages: $\qip(1qfa)$ 
equals $\mathrm{REG}$. This gives a complete characterization of 
the QIP systems with 1qfa verifiers.

\begin{theorem}\label{qfa-regular}
$\mathrm{1QFA}\subsetneqq \qip(1qfa) = \mathrm{REG}$.
\end{theorem}

Note that the first inequality of Theorem \ref{qfa-regular} follows 
from the last equality since $\mathrm{1QFA}\neq \mathrm{REG}$. To 
prove this equality, we first claim in Proposition \ref{regular-1qfa} 
that, for any {\em 1-way deterministic finite 
automaton} ({\em 1dfa}, in short) $M$, we can build a QIP system 
$(P,V)$ in which the 1qfa 
verifier $V$ simulates $M$ in a reversible fashion. Since any move 
of a 1dfa is generally not reversible, 
we need to use an honest prover as an {\lq\lq}eraser{\rq\rq} which 
removes any irreversible information of $M$ into the prover's private 
tape to maintain a history of the verifier's past inner states. This 
simulation establishes the desired inclusion.

\begin{proposition}\label{regular-1qfa} 
$\mathrm{REG}\subseteq \qip_{1,1}(1qfa)$.
\end{proposition}

\begin{proof}
Let $L$ be any regular language and let $M=(Q,\Sigma,\delta_{M})$ be 
any 1dfa that recognizes $L$, where $Q$ is the set of all inner states, 
$\Sigma$ is the input alphabet, and $\delta_{M}$ is the transition 
function. 
We may assume for convenience that $M$'s input tape has the left 
endmarker $\cent$ and the right endmarker $\$$ because this assumption 
does not change the recognition power of the 1dfa. For any pair 
$(q,\sigma)$ of an inner state $q\in Q$ and an input symbol 
$\sigma\in\Sigma$, 
consider the set $S_{q,\sigma}$ of all inner states that lead to $q$ 
while scanning $\sigma$; namely, 
$S_{q,\sigma}=\{p\in Q\mid \delta_{M}(p,\sigma)=q \}$. 

Our goal is to define a QIP system that recognizes $L$ with 
probability $1$. Consider the following QIP protocol that simulates 
$M$ by forcing a prover to act as an eraser. 
In what follows, let 
$\Gamma= \{\#\} \cup \left(\bigcup_{q\in Q,\sigma\in \Sigma} 
S_{q,\sigma}\right)$  be our communication alphabet, provided that  
the symbol $\#$ is not in $Q$. The verifier $V$ is defined to 
simulate truthfully each move of $M$. Let us assume that, at an 
arbitrary step $i\in[1,n+2]_{\integer}$, $V$ is in inner state $p$ 
scanning symbol $\sigma$. 
Now, consider the case where $\delta_{M}(p,\sigma)=q$; 
in other words, $M$ enters state $q$ just after it scans symbol 
$\sigma$ in state $p$. The verifier $V$ behaves as follows. 
In scanning the current communication symbol, whenever it is not 
$\#$, $V$ immediately rejects the input. 
Assuming that the communication symbol is $\#$, $V$ enters the state 
$q$ by passing the communication symbol $p$ to a prover. 
Note that, if the prover always returns $\#$, 
$V$ eventually ends its computation at the time when the head reaches 
the endmarker $\$$. 
If $M$ enters an accepting inner state, then $V$ simply accepts the 
input; otherwise, $V$ rejects the input. 
We design our honest prover $P$ to return $\#$ at every 
communication step.

Let $x$ be any input to our QIP system $(P,V)$. First, consider the 
case where $x$ belongs to $L$. Since the honest prover $P$ erases 
the information on $V$'s inner state at every step, 
$V$ can simulate each move of $M$ in a reversible fashion. 
Hence, $V$ accepts $x$ with probability $1$. On the contrary, when 
$x\not\in L$, 
a dishonest prover $P^*$ cannot return any symbol 
except for $\#$ (or any superposition of such symbols) to optimize 
his adversarial strategy because, otherwise, $V$ can increase his 
rejection probability by immediately entering a rejecting inner 
state in a deterministic manner. 
If $P^*$ always returns $\#$, however, $V$ correctly simulates $M$ 
and eventually enters a rejecting inner state with probability $1$. 
Therefore, $(P,V)$ recognizes $L$ with certainty. 
\end{proof}

To show that $\qip(1qfa)\subseteq \mathrm{REG}$---the opposite 
direction of Proposition \ref{regular-1qfa}, we use two results: 
Lemmas \ref{one-qfa-bound} and \ref{tiling}. To state these lemmas, 
we need the notion of resource-bounded QIP systems. Let $s$ and $t$ 
be any functions mapping $\nat$ to $\nat$. A {\em $(t(n),s(n))$-bounded 
QIP system} is obtained 
from a QIP system by forcing the QIP protocol to {\lq\lq}terminate{\rq\rq} 
after $t(|x|)$ steps on each input $x$ with
$s(|x|)$-space bounded provers. After the $t(|x|)$th measurement, 
we actually stop the entire computation of the QIP system and make 
any non-halting inner state collapse to the special 
output symbol {\lq\lq}{\em I don't know}{\rq\rq}. We say that a 
language {\em $L$ has a $(t(n),s(n))$-bounded 
QIP system} (or {\em a $(t(n),s(n))$-bounded QIP system recognizes 
$L$}) if the system satisfies the completeness and soundness 
conditions given in Section \ref{application} for $L$ with error 
probability at most $\epsilon$, where $\epsilon$ is a certain constant 
drawn from the interval $[0,1/2)$. The following lemma connects 
basic QIP systems to bounded QIP systems.

\begin{lemma}\label{one-qfa-bound}
Let $L$ be any language in $\qip(1qfa)$. There exists a constant 
$c\in\nat^{+}$ such 
that $L$ has an $(n+2,c)$-bounded QIP system with a 1qfa verifier.
\end{lemma}

Lemma \ref{one-qfa-bound} is a direct consequence of Lemma 
\ref{prop2-KM03}, which we shall prove in the subsequent section. 
Another ingredient,  Lemma \ref{tiling}, relates to the notion of 
{\em 1-tiling complexity} \cite{CHPW98}. 
For any language $L$ over alphabet $\Sigma$, we define the infinite 
binary matrix $M_{L}$ whose rows and columns are indexed by 
the strings over $\Sigma$ in the following fashion: any $(x,y)$-entry 
of $M_L$ is $1$ if $xy\in L$ and $0$ otherwise. Furthermore, for each 
$n\in\nat$, $M_L(n)$ denotes the submatrix of $M_L$ whose rows and 
columns are indexed by the strings of length $\leq n$. A {\em 1-tile} 
of $M_L(n)$ is a nonempty submatrix $M$ of $M_{L}(n)$ such that (i) all 
the entries of $M$ are specified by a certain index set $R\times C$, 
where $R,C\subseteq\Sigma^{\leq n}$, and (ii) all the entries of $M$ 
have the same value $1$. For convenience, we often identify $R\times C$ 
with $M$ itself. A {\em 1-tiling} of $M_L(n)$ is a set $S$ of 1-tiles of 
$M_L(n)$ such that every 1-valued entry of $M_L(n)$ is covered by at 
least one element of $S$. The {\em 1-tiling complexity} of $L$ is the 
function $T^1_{L}(n)$ whose value is the minimal size of a 
1-tiling of $M_L(n)$. 

\begin{lemma}\label{tiling}
Let $L$ be any language, let $c\in\nat^+$, and let $\epsilon\in[0,1/2)$. 
If an $(n+2,c)$-bounded QIP system $(P,V)$ with a 1qfa verifier 
recognizes $L$ with error probability at most $\epsilon$, 
then the 1-tiling complexity of $L$ is at most $4^d\ceilings{2\sqrt{2}(1+2d^2)/(1-2\epsilon)}^{2d+1}$, 
where $d$ equals $|Q||\Gamma||\Delta|^c$ for the set $Q$ of the 
verifier's inner states, 
the prover's tape alphabet $\Delta$, and the communication 
alphabet $\Gamma$.
\end{lemma}

\begin{proof} 
Let $L$ be any language recognized by an $(n+2,c)$-bounded QIP 
system $(P,V)$ with a 1qfa verifier 
with error probability at most $\epsilon<1/2$.  
Let $Q$, $\Delta$, and $\Gamma$ be respectively the set of $V$'s inner 
states, $V$'s tape alphabet, 
and the communication alphabet. 
Recall that, for every input $x\in\Sigma^*$ and every step 
$i\in[1,|x|+1]_{\integer}$, 
$U_{P,i}^{x}$ denotes $P$'s $i$th operation on $x$, which is described 
as a $|\Delta|^c$-dimensional unitary matrix 
since $P$ is $c$-space bounded. 
Since $P$'s strategy may differ on a different input, we use the
notation $P_x$ to indicate that 
$P$ always takes the strategy  $\{U^x_{P,i}\}_{i\in\nat^+}$ on any 
given input. 
Write $d$ for $|Q||\Gamma||\Delta|^{c}$ and $\mu$ for 
$(1/2-\epsilon)/(1+2d^2)$. 

Consider the binary matrix $M_L$ induced from $L$. 
Our goal is to present a 1-tiling of $M_{L}(n)$, for each $n\in\nat$, 
of size at most 
$(2\ceilings{\sqrt{2}/\mu})^{2d}\ceilings{1/\mu}\leq 
4^d\ceilings{\sqrt{2}/\mu}^{2d+1} 
=4^d\ceilings{\frac{2\sqrt{2}(1+2d^2)}{1-2\epsilon} }^{2d+1}$. 
Note that, for any 1-valued $(x,y)$-entry of $M_L(n)$, 
since $xy\in L$, the QIP protocol $(P_{xy},V)$ accepts $xy$ with 
probability at least $1-\epsilon$. 
Notationally, for each vector $\bvec{p}$ and any index $\bvec{i}$, 
$[\bvec{p}]_{\bvec{i}}$ represents the $\bvec{i}$-entry of $\bvec{p}$.

For any fixed input $x$, a quadruple $(j_1,j_2,j_3,j_4)$ 
in the set $Q\times [1,|x|+2]_\integer\times \Delta^c\times \Gamma$ 
represents a {\em global configuration} 
of the $(n+2,c)$-bounded QIP system $(P,V)$, 
in which $V$ is in inner state $j_1$ with its head scanning the $j_2$th 
cell, the communication cell contains $j_3$, 
and the prover's private tape consists of $j_4$. 
If the head position $j_2$ is ignored, we call the remaining triplet 
$(j_1,j_3,j_4)$ a {\em semi-configuration}. 
Let $I= Q\times \Delta^c \times\Gamma$ (the set of all 
semi-configurations) 
and, for each $n\in\nat$, let $I_n$ be 
$Q\times [1,n+2]_\integer\times \Delta^c\times\Gamma$ 
(the set of all global configurations on any input of length $n$). 

In the following definition of a 1-tiling, we arbitrarily fix an 
integer $n\in\nat^{+}$ 
and two strings $x$ and $y$ of length $\leq n$ satisfying that $xy\in L$. 
Since $V$ is fixed, we drop the letter $V$ out of  $p_{acc}(x,P,V)$. 
To compute $p_{acc}(xy,P_{xy})$, we introduce two types of vectors. 
The {\em configuration amplitude vector} $\bvec{p}_{x,y}$ is the 
unique $(d+1)$-dimensional vector $\bvec{p}_{x,y}$ 
whose first $d$ entries are indexed by the semi-configurations. 
For simplicity, all the semi-configurations are assumed to be 
enumerated.
For any semi-configuration $\bvec{i}=(i_1,i_2,i_3)\in I$, 
the $\bvec{i}$-entry $[\bvec{p}_{x,y}]_{\bvec{i}}$ is set to be $0$ 
if $i_1$ is a halting inner state; 
otherwise, $[\bvec{p}_{x,y}]_{\bvec{i}}$ is the amplitude of the 
configuration $(i_1,|x|+1,i_2,i_3)$ 
in the superposition obtained after the $|x|+1$st application of 
$V$'s unitary operation. 
In addition, the final $d+1$st entry of $\bvec{p}_{x,y}$ indicates 
the probability 
that $xy$ is accepted within the first $|x|+1$ steps of $V$.  

We further define additional $d$-dimensional vectors to describe the 
transition amplitudes of the protocol $(P,V)$. 
For each index $\bvec{j}=(j_1,j_2,j_3,j_4)\in I_{|y|}$, let 
$\bvec{r}_{x,y}^{\bvec{j}}$ 
be the $d$-dimensional vector whose entries are indexed by the 
semi-configurations $\bvec{i}=(i_1,i_2,i_3)$. 
If $j_1$ is an accepting inner state, then the $(i_1,i_2,i_3)$-entry 
of $\bvec{r}_{x,y}^{\bvec{j}}$ indicates 
the transition amplitude from the configuration $(i_1,|x|+1,i_2,i_3)$ 
to the configuration $(j_1,|x|+j_2+1,j_3,j_4)$; 
otherwise, $[\bvec{r}_{x,y}^{\bvec{j}}]_{\bvec{i}}$ is $0$. 
It immediately follows that $\sum_{\bvec{j}\in I_{|y|} }|[\bvec{r}_{x,y}^{\bvec{j}}]_{\bvec{i}}|^2\leq 1$ 
for any fixed semi-configuration $\bvec{i}\in I$. 

Using the aforementioned vectors, we can calculate the acceptance 
probability $p_{acc}(xy,P_{xy})$ 
of input $xy$ by the protocol $(P_{xy},V)$ as follows: 
$p_{acc}(xy,P_{xy})=\sum_{\bvec{j}\in I_{|y|}}|\bvec{p}'_{x,y}\cdot 
\bvec{r}_{x,y}^{\bvec{j}}|^2
+[\bvec{p}_{x,y}]_{d+1}$, where $\bvec{p}'_{x,y}$ is the 
$d$-dimensional vector 
obtained from $\bvec{p}_{x,y}$ by deleting its last entry and the 
notation $\cdot$ denotes the inner product. 

First, letting $\complex_{1}=\{r\in\complex\mid 
\exists a,b[r= a+ib \;\&\; |a|,|b|\leq 1]\}$, we partition the 
$(d+1)$-dimensional complex space $\complex_{1}^{d}\times [0,1]$ 
into $(2\ceilings{\sqrt{2}/\mu})^{2d}\ceilings{1/\mu}$ 
hyper-cuboids of diameter $\mu$ in each $\complex_{1}$ and $[0,1]$; 
\ie a $\frac{\mu}{\sqrt{2}}\times \frac{\mu}{\sqrt{2}}$ square 
in each of the first $d$ coordinates 
and a real line segment of length $\mu$ in the $d+1$st coordinate. 
Note that some hyper-cuboids near the boundary may have diameter 
less than $\mu$ in certain coordinates. 
Note that each hyper-cuboid has volume at most $\mu^{2d+1}/2^d$. 
Second, we associate each hyper-cuboid $C$ with the rectangle $R_C$ 
defined as 
$R_C=\{x \mid \exists y'(\bvec{p}_{x,y'}\in C\ \wedge \ xy'\in L)\} 
\times\{y\mid \exists x'(\bvec{p}_{x',y}\in C\ \wedge \ x'y\in L)\}$. 
To complete the proof, it suffices to prove that $R_C$ is a 1-tile 
of $M_L(n)$ 
for every hyper-cuboid $C$ whose rectangle is non-empty since, if 
so, every 1-valued entry of $M_L(n)$ is covered by a certain 
1-tile $R_C$ 
and therefore, the collection $T$ of all such rectangles forms a 
1-tiling of $M_L(n)$. 
Hence, the 1-tiling complexity of $L$ is bounded by 
$T^1_L(n) \leq |T| = (2\ceilings{\sqrt{2}/\mu})^{2d}\ceilings{1/\mu}$.  

Let $C$ be any hyper-cuboid whose rectangle is non-empty and let 
$(x,y)$ be any pair of strings of length $\leq n$ in $R_C$. 
Toward a contradiction, we assume that $R_C$ is not a 1-tile; 
namely, $xy\not\in L$. 
This implies that, for any prover $P^*$, $(P^*,V)$ accepts $xy$ with 
probability $\leq \epsilon$. 
Since $(x,y)\in R_C$, there exists a pair $(x',y')$ of strings of 
length $\leq n$ 
such that $\bvec{p}_{x,y'}$ and $\bvec{p}_{x',y}$ are both in $C$. 
It follows that $p_{acc}(x'y,P_{x'y})\geq 1-\epsilon$ since $x'y\in L$. 
Now, consider the special prover $P'$ that simulates $P_{xy'}$ while 
reading $x$ 
and then simulates $P_{x'y}$ while reading $y$. 
By the definition of $P'$, it follows that 
$p_{acc}(xy,P')= \sum_{\bvec{j}\in I_{|y|}}|\bvec{p}'_{x,y'}\cdot
\bvec{r}_{x',y}^{\bvec{j}}|^2 +[\bvec{p}_{x,y'}]_{d+1}$. 

We wish to claim that $p_{acc}(xy,P')>\epsilon$. 
The difference between $p_{acc}(xy,P')$ and $p_{acc}(x'y,P_{x'y})$ 
is upper-bounded by:
\begin{eqnarray*}
\lefteqn{|p_{acc}(xy,P_{xy})-p_{acc}(x'y,P_{x'y})|}\hs{5} \\ 
&\leq&
\left|[\bvec{p}_{x',y}]_{d+1}-[\bvec{p}_{x,y'}]_{d+1}\right|
+
\sum_{\bvec{j}}
\left|
\left|\sum_{\bvec{i}}[\bvec{p}'_{x,y'}]_{\bvec{i}}
[\bvec{r}_{x'y}^{\bvec{j}}]_{\bvec{i}} \right|^2
-\left|\sum_{\bvec{i}}             [\bvec{p}'_{x',y}]_{\bvec{i}}
[\bvec{r}_{x',y}^{\bvec{j}}]_{\bvec{i}} \right|^2 
\right|
\\
&\leq&
\left|[\bvec{p}_{x',y}]_{d+1}-[\bvec{p}_{x,y'}]_{d+1}\right|
+
\sum_{\bvec{j}}
\sum_{\bvec{i},\bvec{i'}}\left|
([\bvec{p}'_{x,y'}]_{\bvec{i}}[\bvec{p}'_{x,y'}]^*_{\bvec{i'}}
-[\bvec{p}'_{x',y}]_{\bvec{i}}[\bvec{p}'_{x',y}]^*_{\bvec{i'}})
[\bvec{r}_{x',y}^{\bvec{j}}]_{\bvec{i}} 
[\bvec{r}_{x',y}^{\bvec{j}}]^*_{\bvec{i'}} 
\right|.
\end{eqnarray*}
The first term 
$\left|[\bvec{p}_{x',y}]_{d+1}-[\bvec{p}_{x,y'}]_{d+1}\right|$ 
is at most $\mu$ since $\bvec{p}_{x',y}$ 
and $\bvec{p}_{x,y'}$ are in the same hyper-cuboid.
The last term is also bounded above by 
$2\mu \sum_{\bvec{j}}\sum_{\bvec{i},\bvec{i'}}
\left|[\bvec{r}_{x',y}^{\bvec{j}}]_{\bvec{i}} 
[\bvec{r}_{x',y}^{\bvec{j}}]^*_{\bvec{i'}}\right|$. 
This comes from the following bound: 
\[
|[\bvec{p}'_{x,y'}]_{\bvec{i}}[\bvec{p}'_{x,y'}]^*_{\bvec{i'}}
-[\bvec{p}'_{x',y}]_{\bvec{i}}[\bvec{p}'_{x',y}]^*_{\bvec{i'}}|
\leq 
|[\bvec{p}'_{x,y'}]_{\bvec{i}}([\bvec{p}'_{x,y'}]^*_{\bvec{i'}}-
[\bvec{p}'_{x',y}]^*_{\bvec{i'}})| + 
|[\bvec{p}'_{x',y}]^*_{\bvec{i'}}([\bvec{p}'_{x,y'}]_{\bvec{i}}-
[\bvec{p}'_{x',y}]_{\bvec{i}})| 
\leq 2\mu.
\]
This term $2\mu \sum_{\bvec{j}}\sum_{\bvec{i},\bvec{i'}}
\left|[\bvec{r}_{x',y}^{\bvec{j}}]_{\bvec{i}} 
[\bvec{r}_{x',y}^{\bvec{j}}]^*_{\bvec{i'}}\right|$ 
is further bounded by $2\mu \sum_{\bvec{i},\bvec{i'}}
\sqrt{\sum_{\bvec{j}}|[\bvec{r}_{x',y}^{\bvec{j}} ]_{\bvec{i}}|^2 }
\sqrt{\sum_{\bvec{j}}|[\bvec{r}_{x',y}^{\bvec{j}} ]^*_{\bvec{i'}}|^2 }$ 
using the Cauchy-Schwarz inequality and is thus at most $2\mu d^2$.
Overall, the term $|p_{acc}(xy,P')-p_{acc}(x'y,P_{x'y})|$ is 
upper-bounded by $\mu(1+ 2d^2)$, which also equals $1/2-\epsilon$ 
by the choice of $\mu$.
Therefore, the desired inequality $p_{acc}(xy,P')\geq 1/2>\epsilon$ 
follows immediately from $p_{acc}(x'y,P_{x'y})\geq 1-\epsilon$. 
This implies that $(P',V)$ accepts $xy$ with probability $>\epsilon$. 
This contradicts our assumption that $p_{acc}(xy,P^*)\leq \epsilon$ 
for any prover $P^*$. Therefore, $R_C$ is a 1-tile of $M_L(n)$. 
\end{proof}

At length, we obtain the containment $\qip(1qfa)\subseteq \mathrm{REG}$ 
by combining Lemmas \ref{one-qfa-bound} and \ref{tiling}, 
which indicates that every language in $\qip(1qfa)$ has 1-tiling 
complexity $O(1)$. 
Recall from \cite{CHPW98} that a language is regular if and only 
if its 1-tiling complexity is bounded 
above by a certain constant. Therefore, it immediately follows that 
$\qip(1qfa)\subseteq\mathrm{REG}$, as requested. This completes 
the proof of Theorem \ref{qfa-regular}.

\section{Two-Way QFA Verifiers against Mighty Provers}

We have seen in the previous section that, using interactions 
with provers, 
1qfa verifiers can exercise a remarkable power of recognizing the 
regular languages. 
This section turns our interest to the 2qfa-verifier QIP systems; 
namely, $\qip(2qfa)$ and $\qip(2qfa,poly\mbox{-}time)$. 
First, observe that a verifier can completely eliminate any 
intrusion of a prover by simply ignoring the communication cell 
(\ie applying the identity operation). This observation yields 
the following simple containments: 
$\mathrm{2QFA}\subseteq \qip(2qfa)$ and 
$\mathrm{2QFA}(poly\mbox{-}time)\subseteq 
\qip(2qfa,poly\mbox{-}time)$. 

Now, we demonstrate the power of 
$\qip(2qfa,poly\mbox{-}time)$.

\begin{theorem}\label{am-polytime}
$\mathrm{REG}\subsetneqq \qip(2qfa,poly\mbox{-}time) 
\nsubseteq \am(2pfa)$. 
\end{theorem}

The first proper containment follows immediately from the 
facts that $\mathrm{REG}\subsetneqq \mathrm{2QFA}(poly\mbox{-}time)$ 
\cite{KW97} and $\mathrm{2QFA}(poly\mbox{-}time)\subseteq 
\qip(2qfa,poly\mbox{-}time)$. To prove the second separation, 
we first introduce a variation of $Pal$, briefly called $Pal_{\#}$, 
which is defined as $Pal_{\#}=\{x\#x^{R}\mid x\in\{0,1\}^*\}$ 
over the alphabet $\{0,1,\#\}$, 
where $\#$ is a separator not in $\{0,1\}$. Similar to $Pal$ 
\cite[Theorem 3.4]{DS92}, we can show that this language 
$Pal_{\#}$ does not belong to $\am(2pfa)$. In the following 
lemma, we further claim that $Pal_{\#}$ is indeed in 
$\qip(2qfa,poly\mbox{-}time)$. Theorem \ref{am-polytime} 
naturally follows from this lemma. 

\begin{lemma}\label{pal-sharp}
For any constant $\epsilon\in(0,1/2]$, $Pal_{\#}\in \qip_{1,1-\epsilon}(2qfa,poly\mbox{-}time)$.
\end{lemma}

\begin{proof}
We slightly modify the classical IP protocol given in 
\cite{DS92} for $Pal$. Let $\Sigma=\{0,1,\#\}$ be our input 
alphabet, let $\Gamma =\{0,1,\#\}$ be our communication alphabet.
Let $\epsilon$ be any error bound in $(0,1/2]$ and set 
$d=\ceilings{\log_2(1/\epsilon)}$. Note that $d\geq1$ since 
$\epsilon\leq 1/2$. Our QIP system $(P,V)$ for $Pal_{\#}$ is 
given as follows. We begin with the description of the 2qfa 
verifier $V$ who runs in worst-case linear time. 
Recall that the verifier's head is initially scanning the 
endmarker $\cent$ with the blank symbol $\#$ in the communication 
cell. Let $x$ be any input string. 
The verifier runs the following quantum algorithm by stages, 
creating the total of $2^d$ independent computation paths. 
The initial stage is assumed to be $s=\lambda$, the empty string.
\begin{quote}
Repeat the following procedure (*) until $|s|=d$. 
During this procedure, $V$ always unalters the communication cell 
(such a verifier is said to make a {\em one-way communication}). 
Assume that $V$ is in stage $s\in\{0,1\}^{\leq d-1}$.

(*) In the first phase, the head moves rightward. If there 
is no $\#$ in $x$, 
then $V$ rejects $x$ when $V$ scans the right endmarker. In 
scanning $\#$, 
$V$ generates a superposition of two independent branches by 
entering two inner states $q_{1,s}$ and $q_{2,s}$ 
with the equal amplitude $1/\sqrt{2}$. In the branch starting 
with $q_{1,s}$, the head moves leftward; 
in the other branch with $q_{2,s}$, it moves rightward. During 
this phase, 
whenever a prover returns any non-blank symbol, $V$ rejects $x$ 
immediately. 
In the second phase, visiting each cell, $V$ receives a communication 
symbol, say $a$, from a prover. 
The head checks whether it is currently scanning $a$ in the input 
tape unless the head arrives at an endmarker. 
If $V$ discovers a discrepancy, then it enters a rejecting inner 
state. 
When the head reaches an endmarker, $V$ rejects $x$ if a prover 
sends a non-blank symbol. 
The head at the left endmarker $\cent$ stays still for another 
step and enters $q_{0,s0}$ 
whereas the head at the right endmarker $\$$ enters $q_{0,s1}$ 
by moving right to $\cent$ 
(since the input tape is circular\footnote{The circularity of 
the input tape is used to simplify the description of the 
transitions and is not necessary for the lemma.}). Go to 
the next stage.

Along each computation path $s$, if $x$ is not yet rejected 
after executing (*) $d$ times, 
then $V$ enters an accepting inner state.
\end{quote}
Table \ref{table:pal-sharp} describes the formal transitions of $V$. 
Note that the running time of $V$ is $O(n)$ even in the worst case.  
Consider the case where $x=y\#y^{R}$ for a certain string $y$. 
In each round, the honest prover $P$ must pass the string $y^{R}$ 
bit by bit to the verifier after $V$ splits into two branches. 
With this honest prover $P$, $V$ never enters any rejecting inner 
state. Hence, after $d$ rounds, $V$ finally accepts $x$ with 
probability $1$.

\begin{table}[ht]
\bs\begin{center}
\begin{tabular}{|ll|}\hline
$V_{\cent}\qubit{q_{0,s}}\qubit{\#}=\qubit{q'_{0,s}}\qubit{\#}$ 
& $V_{\cent}\qubit{q_{1,s}}\qubit{\#}=\qubit{q_{0,s0}}\qubit{\#}$ \\ 
$V_{\cent}\qubit{q_{i,s}}\qubit{a}=\qubit{r_{i,s}}\qubit{a}$ 
& $V_{\$}\qubit{q_{i,s}}\qubit{a}=\qubit{r_{i,s}}\qubit{a}$ \\
$V_{\$}\qubit{q'_{0,s}}\qubit{\#} =\qubit{r_{0,s}}\qubit{\#}$ 
& $V_{\$}\qubit{q_{2,s}}\qubit{\#} =\qubit{q_{0,s1}}\qubit{\#}$ \\ 
$V_{\#}\qubit{q'_{0,s}}\qubit{\#}= 
\frac{1}{\sqrt{2}}(\qubit{q_{1,s}}\qubit{\#} + \qubit{q_{2,s}}\qubit{\#})$ 
& $V_{\#}\qubit{q_{i,s}}\qubit{b}=\qubit{r_{i,s}}\qubit{b}$ \\
$V_{a}\qubit{q'_{0,s}}\qubit{\#} = \qubit{q'_{0,s}}\qubit{\#}$ & \\
$V_{a}\qubit{q_{i,s}}\qubit{a} =\qubit{q_{i,s}}\qubit{a}$ &  
$V_{a}\qubit{q_{i,s}}\qubit{a'}= \qubit{r_{i,s}}\qubit{a'}$  \\
& \\ 
$D(q_{1,s})=-1$  & $D(q_{2,s})=1$ \\
$D(q'_{0,s})=1$ & $D(q_{0,s0})=0$ \\
$D(q_{0,s1})=1$ & \\  \hline
\end{tabular}
\caption{Transitions of $V$ for $Pal_{\#}$ with stage 
$s\in\{0,1\}^{\leq d-1}$, $i\in\{1,2\}$, 
$a\in\{0,1\}$, $a'\in\Gamma$ with $a\neq a'$, and $b\in\Gamma$. 
The rejecting inner states are $r_{0,s}$ and $r_{i,s}$. 
The unitary operator $V_{\sigma}$ for each $\sigma\in\check{\Sigma}$ 
acts on the Hilbert space spanned by $Q\times\Gamma$. 
The transition function $\delta$ of $V$ is then induced by setting 
$\delta(q,\sigma,\gamma,q',\gamma',d)=
\langle q',\gamma'|U_\sigma|q,\gamma\rangle$ 
and $d=D(q)$ for any $q,q'\in Q$ and any 
$\gamma,\gamma'\in\Gamma$. 
}\label{table:pal-sharp}  
\end{center}
\end{table}

Next, assume that $x$ is not in $Pal_{\#}$. 
It suffices to consider only the case where $x$ is of the 
form $y\#z^{R}$ since, 
if there is no $\#$, $V$ rejects $x$ with probability $1$. 
In each round, since $V$ makes only one-way communication 
with a dishonest prover, 
the prover's visible configuration is exactly the same 
along two branches. 
In other words, the prover answers in exactly the same way 
along these two branches. 
In the second phase, a dishonest prover $P^*$ may return a 
superposition of $0$ and $1$. 
Since $V$'s two branches never interfere with each other in 
each round, 
$V$ can eliminate at least one of them by entering a rejecting 
inner state. 
This gives the rejection probability of at least $1/2$ 
since the squared magnitude of the superposition obtained along 
each branch is exactly $1/2$. 
Since we repeat (*) $d$ times, the total rejection probability 
sums up to at least $\sum_{i=1}^{d}2^{-i} = 1-1/2^d$, 
which is lower-bounded by $1-\epsilon$ by the choice of $d$. 
Thus, $V$ rejects $x$ with probability $\geq 1-\epsilon$. 
Therefore, $(P,V)$ is a $(1,1-\epsilon)$-QIP system that 
recognizes $Pal_{\#}$.
\end{proof}

Supplementing Theorem \ref{am-polytime}, we now present an 
upper bound of $\qip(2qfa,poly\mbox{-}time)$: with an 
appropriate choice of amplitudes, 
$\qip(2qfa,poly\mbox{-}time)$ is located in the complexity 
class  $\np$, 
where $\np$ is the class consisting of all languages recognized 
by nondeterministic Turing machines in polynomial time.  
This can be compared with a result of Dwork and Stockmeyer \cite{DS92}, 
who proved that 
$\am(2pfa) \subsetneqq \ip(2pfa,poly\mbox{-}time)\subseteq \pspace$. 

\begin{theorem}\label{qip-vs-np}
$\qip_{\tilde{\complex}}(2qfa,poly\mbox{-}time) \subseteq \np$.
\end{theorem}

To show the desired upper-bound of 
$\qip_{\tilde{\complex}}(2qfa,poly\mbox{-}time)$, 
we need the following lemma, which is similar to Lemma 
\ref{one-qfa-bound}.
 
\begin{lemma}\label{one-qfa-size}
Every language in $\qip(2qfa,poly\mbox{-}time)$ has a 
$(t(n),c\log{n}+c)$-bounded QIP system 
for a certain polynomial $t$ and a certain constant $c>0$.
\end{lemma}

Lemma \ref{one-qfa-size} (as well as Lemma \ref{one-qfa-bound}) 
directly comes from the following lemma 
whose proof is based on the result of Kobayashi and Matsumoto 
\cite{KM03}. Lemma \ref{prop2-KM03} states that, without changing 
the acceptance probability, the prover's visible configuration 
space can be reduced in size to the verifier's visible 
configuration space. 

\begin{lemma}\label{prop2-KM03} 
Let $(P,V)$ be any QIP system with a 2qfa (1qfa, resp.) verifier 
and let $Q,\Gamma$ be 
respectively the sets of all inner states and of all communication 
symbols. 
There exists another prover $P'$ that satisfies the following 
two conditions: for every 
input $x$ and every $i\in\nat^+$, (i) the prover's $i$th 
operation $U^x_{P',i}$ is a $|Q||\Gamma|(|x|+2)$-dimensional 
($|Q||\Gamma|$-dimensional, resp.) unitary operator, and 
(ii) $(P',V)$ accepts $x$ with the same probability as $(P,V)$ 
does. 
\end{lemma}

\begin{proof}
Take an arbitrary QIP system $(P,V)$ with the set $Q$ of all 
inner states of $V$, the communication alphabet $\Gamma$, 
and the transition function $\delta$ of $V$. 
In this proof, we consider only the case where $V$ is a 1qfa. 
The remaining case where $V$ is a 2qfa can be similarly proven 
if we further include the information on $V$'s head position. 

For convenience, we view our QIP system $(P,V)$ as a quantum 
circuit of three registers. 
The first register represents the inner state of $V$ together 
with the head position of the input tape, 
the second register represents the communication cell, 
and the third register represents a prover's private tape. 
Let $x$ be any input of length $n$. 
Recall the Hilbert spaces $\VV_n$, $\MM$, and $\PP$ associated 
with $(P,V)$ on input $x$. 
The Hilbert space $\VV_n$ is the tensor product of the 
$|Q|$-dimensional space $\VV$ 
and the $(n+2)$-dimensional space $\VV'_n$. 
Henceforth, we can omit the description of qubits on $\VV'_n$ 
since $V$ is a 1qfa.  
The initial superposition of $(P,V)$ is 
$\qubit{\chi_0}=\qubit{q_0}\qubit{\#}\qubit{\lambda}$. 
Note that, at each step $i\in[1,n+1]_{\integer}$, without changing 
$V$'s acceptance probability, 
we can swap the application order of $V$'s $i$th measurement 
$E_{non}$ and $P$'s $i$th operation $U_{P,i}^{x}$.
For each index $i\in[1,n+1]_{\integer}$, the three superpositions 
$\qubit{\phi_i}$, $\qubit{\psi_i}$, and $\qubit{\chi_i}$ 
are inductively defined as follows:  
$\qubit{\chi_i} = E_{non}\qubit{\psi_i}$,  $\qubit{\psi_i} = 
U^x_{P,i}\qubit{\phi_i}$, and  
$\qubit{\phi_i} = U_\delta^x \qubit{\chi_{i-1}}$.
In addition, let $\qubit{\phi_{n+2}}= U^x_{\delta}\qubit{\chi_{n+1}}$, 
which is the superposition obtained just before the final 
measurement. 

For brevity, write $\PP'$ for the $|Q||\Gamma|$-dimensional 
Hilbert space 
that corresponds to the private tape of a $|Q||\Gamma|$-space 
bounded prover. 
Our goal is to define the prover $P'$ that works on $\MM\otimes\PP'$. 
Hereafter, we define the strategy $\{U^x_{P',i}\}_{i\in\nat^{+}}$ 
of $P'$ on input $x$. 
For convenience, set $\qubit{\chi'_0}=\qubit{\chi_0}$. 
It follows from \cite[page 110]{NC00} that, for every index 
$i\in[1,n+1]_{\integer}$, 
there exists a vector $\qubit{\psi'_i}$ in $\VV\otimes\MM\otimes\PP'$ 
satisfying that 
$\mathrm{tr}_{\PP'}\ketbra{\psi'_i}{\psi'_i}= 
\mathrm{tr}_{\PP}\ketbra{\psi_i}{\psi_i}$ 
since the dimension of $\PP'$ is the same as that of 
$\VV\otimes\MM$. 
We further define the vectors $\qubit{\chi'_i}$ and 
$\qubit{\phi'_i}$ as follows: 
let $\qubit{\chi'_i}=E_{non}\qubit{\psi'_i}$ for 
$i\in[1,n+1]_{\integer}$ 
and $\qubit{\phi_i'}=U^{x}_{\delta}\qubit{\chi_{i-1}'}$ for 
any $i\in[1,n+2]_{\integer}$. Note that 
$\mathrm{tr}_{\PP}\ketbra{\phi_1}{\phi_1}=
\mathrm{tr}_{\PP'}\ketbra{\phi'_1}{\phi'_1}$. 
Now, fix $j\in[2,n+2]_{\integer}$ arbitrarily. 
Since $U^{x}_{\delta}$ and $E_{non}$ act on neither 
$\PP$ nor $\PP'$, we obtain: 
\[
\mathrm{tr}_{\PP}\ketbra{\phi_j}{\phi_j}= 
U^{x}_{\delta}E_{non}(\mathrm{tr}_{\PP}\ketbra{\psi_{j-1}}{\psi_{j-1}})
E_{non}(U^{x}_{\delta})^\dagger 
= U^{x}_{\delta}E_{non}(\mathrm{tr}_{\PP'}\ketbra{\psi_{j-1}'}{\psi_{j-1}'})
E_{non}(U^{x}_{\delta})^\dagger 
= \mathrm{tr}_{\PP'}\ketbra{\phi_j'}{\phi_j'},
\]
which further implies: for any $i\in[1,n+1]_{\integer}$,
\[
\mathrm{tr}_{\MM\otimes\PP'}\ketbra{\psi'_i}{\psi'_i}
=\mathrm{tr}_{\MM\otimes\PP}\ketbra{\psi_i}{\psi_i}
=\mathrm{tr}_{\MM\otimes\PP}\ketbra{\phi_i}{\phi_i} 
=\mathrm{tr}_{\MM\otimes\PP'}\ketbra{\phi'_i}{\phi'_i},
\]
where the second equality comes from the fact that 
$U^x_{P,i}$ is applied only to the space ${\cal M}\otimes{\cal P}$. 
Since $\mathrm{tr}_{\MM\otimes\PP'}\ketbra{\psi'_i}{\psi'_i}
=\mathrm{tr}_{\MM\otimes\PP'}\ketbra{\phi'_i}{\phi'_i}$,
there exists a unitary operator $U_i$ acting on $\MM\otimes\PP'$ 
satisfying that $(I\otimes U_i)\qubit{\phi'_i}=\qubit{\psi'_i}$ 
\cite{HJW93,Uhl86}. 
The desired operation $U_{P',i}^{x}$ of $P'$ is set to be this 
$I\otimes U_i$. 

Next, we compare the acceptance probabilities of the two QIP 
systems $(P,V)$ and $(P',V)$. 
We have $\mathrm{tr}_{\PP}\ketbra{\psi_i}{\psi_i} 
=\mathrm{tr}_{\PP'}\ketbra{\psi'_i}{\psi'_i}$ 
for every $i\in[1,n+1]_\integer$ as well as 
$\mathrm{tr}_{\PP}\ketbra{\phi_{n+2}}{\phi_{n+2}}
= \mathrm{tr}_{\PP'}\ketbra{\phi'_{n+2}}{\phi'_{n+2}}$. 
Thus, for every $i\in[1,n+2]_\integer$, the acceptance 
probability of $x$ produced by the $i$th measurement of $(P,V)$ 
equals the acceptance probability of $x$ caused by the $i$th 
measurement of $(P',V)$. 
This completes the proof.  
\end{proof}

Lemma \ref{prop2-KM03} lets us focus our attention only on 
$(n^{O(1)},O(\log(n)))$-bounded QIP systems. 
To simulate such a system, we need to approximate the prover's unitary 
operations using only a fixed universal set of quantum gates. Lemma 
\ref{qustring-gate} relates to an upper bound of the number of quantum 
gates necessary to approximate a given unitary operator. The lemma, 
explicitly stated in \cite{NY04}, can be obtained from the 
Solovay-Kitaev theorem \cite{Kit97,NC00} 
following the standard decomposition of unitary matrices. We 
fix an appropriate universal set  of quantum gates
consisting of the Controlled-NOT gate and a finite number of single-qubit 
gates, with $\tilde{\complex}$-amplitudes, that generate a dense subset 
of SU(2) with their inverse. Write $\log^k{n}$ for $(\log{n})^k$ 
for any constant $k\in\nat^{+}$.

\begin{lemma}\label{qustring-gate} 
For any sufficiently large $k\in\nat^{+}$, any $k$-qubit unitary 
operator $U_k$, and any real number $\epsilon>0$, there 
exists a quantum circuit 
$C$ of size at most 
$2^{3k}\log^3{(1/\epsilon)}$ acting on $k$ qubits such that  
$\|U_{C}-U_k\|<\epsilon$, where $U_{C}$ 
is the unitary operator corresponding to $C$, where $\|A\|= 
\sup_{\qubit{\phi}\neq0}\|A\qubit{\phi}\|/\|\qubit{\phi}\|$.
\end{lemma}

A quantum circuit $C$ built in Lemma \ref{qustring-gate} can 
be further 
encoded into a binary string, provided that the encoding length 
is at least the size of the quantum circuit. 
This enables us to prove the simulation result of any bounded QIP 
system with $\tilde{\complex}$-amplitudes. 
We say that a function $f$ from $\nat$ to $\nat$ is {\em 
polynomial-time computable} 
if there exists a deterministic Turing machine that, on any 
input $1^n$, outputs $1^{f(n)}$. 

\begin{proposition}\label{NTIME}
Let $s$ and $t$ be any polynomial-time computable functions from 
$\nat$ to $\nat$. 
Any language that has a $(t(n),s(n))$-bounded QIP system with a 
2qfa verifier using 
$\tilde{\complex}$-amplitudes belongs to the complexity class $\mathrm{NTIME}(n^{O(1)}t(n)2^{O(s(n))}\log^{O(1)}t(n))$.
\end{proposition}

The proof of Proposition \ref{NTIME} is outlined as follows. 
Given a bounded QIP system, we first guess a binary string that 
encodes a quantum circuit 
representing the prover's strategy. 
We then simulate the verifier's move followed by the prover's 
operation. 
This simulation can be done deterministically by listing all 
the verifier's configurations and 
simulating their amplitudes at each step. 
After each step of the verifier, we calculate the probability 
of reaching any halting configuration instead 
of performing measurement. 
Now, we give the formal proof of Proposition \ref{NTIME}.

\begin{proofof}{Proposition \ref{NTIME}} 
Let $(P,V)$ be any $(t(n),s(n))$-bounded QIP system with a 
2qfa verifier and 
let $A$ be the language recognized by $(P,V)$ with error 
probability at most $1/2-\epsilon$ 
for a certain fixed constant $\epsilon\in(0,1/2]$. 
Let $x$ be any input string of length $n$. 
By translating the prover's tape alphabet $\Delta$ to 
$\{0,1\}^{\ceilings{\log|\Delta|}}$ 
and the communication alphabet $\Gamma$ to 
$\{0,1\}^{\ceilings{\log|\Gamma|}}$, 
we can assume without loss of generality 
that our prover uses at most $\ceilings{\log|\Delta|}s(n)$ 
qubits on  his private tape 
and writes $\ceilings{\log|\Gamma|}$-qubit strings in the 
communication cell. 
Now, let $s'(n)= \ceilings{\log|\Delta|}s(n)+\ceilings{\log|\Gamma|}$ 
for any $n\in\nat$.

A prover comprises a series of $t(n)$ unitary matrices on $s'(n)$ 
qubits, say $U_1,U_2,\ldots,U_{t(n)}$. 
For each $U$ of such matrices, Lemma \ref{qustring-gate} gives a 
quantum circuit $C_{U}$ 
of size at most $2^{3s'(n)}\log^3(d t(n))$ such that 
the unitary operator associated with $C_{U}$ approximates $U$ to 
within $1/dt(n)$, 
where $d$ is a constant satisfying $d>2/\epsilon$.  
This makes it possible to replace the prover $P$ by the series of 
$t(n)$ quantum circuits 
$(C_{U_1},C_{U_2},\ldots,C_{U_{t(n)}})$, which is hereafter abbreviated $C$. 
Note that the cumulative approximation error is bounded above by 
$\sum_{i=1}^{t(n)}\frac{1}{dt(n)} =1/d$, 
which is smaller than $\epsilon/2$. 
Using this $C$ as a prover, $V$ proceeds his computation and accepts 
(rejects, resp.) $x$ 
with probability $\geq(1/2+\epsilon)-\epsilon/2 = 1/2+ \epsilon/2$ 
if $x\in A$ ($x\not\in A$, resp.). 
Choose an effective encoding $\pair{C}$ of $C$ satisfying that 
$|\pair{C}|\leq ct(n)\cdot 2^{3s'(n)}\log^3(dt(n))$ 
for a certain constant $c>0$. 
Note that any configuration of $(C,V)$ requires $s'(n)+O(\log n)$ qubits. 

Using the encoding $\pair{C}$, we give a classical simulation of the 
computation of $(C,V)$ on input $x$. 
Note that the verifier $V$ can be represented by the product of 
$t(n)+1$ unitary matrices of dimension polynomial in $n$ 
and the {\lq\lq}prover{\rq\rq} $C$ consists of $t(n)$ unitary 
matrices of dimension $2^{s'(n)}$. 
Note that all the gates in $C$ and verifier's transition function 
use only polynomial-time approximable amplitudes. 
Within time polynomial in $n$ and $\log t(n)$, we can approximate 
such amplitudes to within $\frac{1}{t(n)2^{r(n)}}$ 
for any fixed polynomial $r$.   
By choosing a sufficiently large polynomial $r$, we can 
deterministically simulate with high accuracy the computation 
of $(C,V)$ in polynomial time. 
Such a simulation gives an approximation of the acceptance 
probability $p_{acc}(x,C,V)$. 
Now, we accept the input $x$ if the approximated acceptance 
probability exceeds $1/2$, and reject $x$ otherwise. 
For a certain polynomial $p$ independent of $n$, we therefore 
obtain a $t(n)2^{O(s(n))}p(n,\log t(n))$-time 
deterministic algorithm that approximately simulates $V$ 
with a fixed prover $C$. 

At last, we consider the following nondeterministic algorithm $\AAA$: 
\begin{quote}
On input $x$ ($n=|x|$), nondeterministically guess $\pair{C}$, 
where $C$ is a series of $t(n)$ quantum circuits 
of size $\leq 2^{3s'(n)}\log^3(dt(n))$. 
If the aforementioned deterministic simulation of $(C,V)$ leads 
to acceptance, then accept $x$, or else reject $x$.  
\end{quote}
It is easy to verify that $\AAA$ recognizes $L$ 
in time $p'(n,\log t(n))\cdot 2^{O(s(n))}t(n)$ for an appropriate  
polynomial $p'$. 
Therefore, $L$ belongs to 
$\mathrm{NTIME}(n^{O(1)}t(n)2^{O(s(n))}\log^{O(1)} t(n))$.
\end{proofof}

We return to the proof of the second part of Theorem 
\ref{qip-vs-np}. Take any language $L$ in 
$\qip_{\tilde{\complex}}(2qfa,poly\mbox{-}time)$. Lemma 
\ref{one-qfa-size} guarantees the existence of a bounded-error 
$(t(n),s(n))$-bounded QIP system recognizing $L$ using 
$\tilde{\complex}$-amplitudes, where $t(n)$ is a polynomial 
and $s(n)$ is a logarithmic function. 
{}From Proposition \ref{NTIME}, it follows that $L$ belongs 
to the complexity class 
$\mathrm{NTIME}(n^{O(1)}t(n)2^{O(s(n))}\log^{O(1)} t(n))$, 
which clearly coincides with $\np$. This ends the proof of Theorem 
\ref{qip-vs-np}.

In the end of this section, we present a closure property of 
QIP systems with 2qfa verifiers.

\begin{proposition}\label{union}
$\qip(2qfa)$ and $\qip(2qfa,poly\mbox{-}time)$ are closed under union.
\end{proposition}

Proposition \ref{union} is shown in the following fashion. 
For any two 2qfa-verifier QIP systems 
$(P_1,V_1)$ and $(P_2,V_2)$ that respectively correspond to 
$L_1$ and $L_2$, the verifier for $L_1\cup L_2$ 
first asks a prover to choose the minimal index $i\in\{1,2\}$ 
for which $(P_i,V_i)$ 
accepts $x$ (if $i$ exists). The verifier then simulates the 
protocol $(P_i,V_i)$ to check 
whether $(P_i,V_i)$ truly accepts $x$. The formal proof below 
shows the validity of this protocol.

\begin{proofof}{Proposition \ref{union}}
We prove only the closure property of $\qip(2qfa)$ under union because a similar proof shows the closure property of $\qip(2qfa,poly\mbox{-}time)$. 
Take any two languages $L_1,L_2\in\qip(2qfa)$ and, for each $i\in\{1,2\}$, choose a QIP system $(P_i,V_i)$ that recognizes $L_i$ with error probability $\leq \epsilon$, where $\epsilon$ is any fixed constant in $[0,1/2)$. Without loss of generality, we may assume that the set of all inner states of $V_1$ and that of $V_2$ are mutually disjoint. 
Consider the following protocol of a new verifier $V$ to determine whether any given input $x$ belongs to $L_1\cup L_2$. 
At the first move, $V$ sends the communication symbol $\#$ to a prover without moving its tape head and waits for the prover's reply $i\in\{1,2\}$. 
Whenever the reply $i$ is neither $1$ nor $2$, $V$ immediately rejects $x$ to prevent the prover from tampering. On the contrary, if $i$ is truly in $\{1,2\}$, then $V$ simulates $V_i$. On any input $x$ in $L_1\cup L_2$, our honest prover $P$ first returns the minimal index $i\in\{1,2\}$ such that $x\in L_i$ and then behaves like $P_i$.

Henceforth, we prove that $(P,V)$ recognizes $L_1\cup L_2$. Let $x$ be an arbitrary input. First, assume that $x\in L_1\cup L_2$. Obviously, if $x\in L_1$, then the protocol $(P,V)$ simulates $(P_1,V_1)$ and otherwise, $(P,V)$ simulates $(P_2,V_2)$. Hence, $V$ accepts $x$ with probability at least $1-\epsilon$. 
Next, assume that $x\not\in L_1\cup L_2$. To maximize the acceptance probability of $V$ on the input $x$, 
a dishonest prover should return either $1$ or $2$ (or their superposition). However, $V$ simulates $V_i$ 
when he receives $i$, and the computation paths of $V$ that simulate $V_1$ and $V_2$ do not interfere 
with each other. Thus, for any prover $P^*$, $(P^*,V)$ rejects $x$ with probability at least $1-\epsilon$.  
This completes the proof. 
\end{proofof}

\section{How Often is Measurement Performed?}

Measurement is one of the most fundamental operations in quantum computation. Although a measurement is necessary to {\lq\lq}know{\rq\rq} the content 
of a target quantum state, the measurement collapses the quantum state 
and thus causes a quantum computation irreversible. Since a qfa uses only a finite amount of memory space, the number of times when measurements are conducted affects the computational power in general. Recall measure-once 1qfa's or mo-1qfa's from Section \ref{sec:QFA}. 
We define an {\em mo-1qfa verifier} as a 1qfa verifier who does not perform any measurement until he applies the final unitary operation while visiting the right endmarker $\$$. This indicates that a measurement takes place only once after the verifier makes exactly $|x|+2$ moves on input $x$. We use the restriction $\pair{mo\mbox{-}1qfa}$ to indicate 
that a verifier is an mo-1qfa. This section makes a comparison 
between mo-1qfa verifiers and 1qfa verifiers in our QIP systems. As mentioned in 
Section \ref{sec:QFA}, mo-1qfa's and 1qfa's are quite different in power 
because of the different numbers of measurement operations performed during a computation. 

In what follows, we show that (i) the QIP systems with mo-1qfa verifiers 
are more powerful than mo-1qfa's alone and (ii) mo-1qfa verifiers are more 
prone to be fooled by dishonest provers than 1qfa verifiers. 
 
\begin{theorem}\label{mo-1qfa-regular}
$\mathrm{MO\mbox{-}1QFA}\subsetneqq 
\qip(mo\mbox{-}1qfa) \subsetneqq \qip(1qfa)$. 
\end{theorem}

Theorem \ref{mo-1qfa-regular} is a direct consequence of Proposition 
\ref{mo-complement}, which refers to a closure property of $\qip(mo\mbox{-}1qfa)$. 
Conventionally, a complexity class $\CC$ is said to be {\em closed under complementation} 
if, for any language $A$ over alphabet $\Sigma$ in $\CC$, its complement $\Sigma^*- A$ is also in $\CC$. 

\begin{proposition}\label{mo-complement}
$\qip(mo\mbox{-}1qfa)$ is not closed under complementation.
\end{proposition}

Theorem \ref{mo-1qfa-regular} follows from Proposition 
\ref{mo-complement} because $\qip(1qfa)$ ($=\mathrm{REG}$) and $\mathrm{MO\mbox{-}1QFA}$ 
are known to be closed under complementation \cite{MC00}. 

To prove Proposition \ref{mo-complement}, it suffices to show that (i) the unary language $L_a= \{a\}^*-\{\lambda\}$ is in $\qip_{1,1}(mo\mbox{-}1qfa)$ and (ii) the language  $\{\lambda\}$ is not in $\qip(mo\mbox{-}1qfa)$. 
We first show that $L_a\in \qip_{1,1}(mo\mbox{-}1qfa)$. 
We set out alphabets $\Sigma$ and $\Gamma$ as $\Sigma=\{a\}$ and $\Gamma=\{a,\#\}$. 
The transition of our verifier $V$ is given in Table \ref{table:La}. 
At the first step, $V$ stays in the initial inner state $q_0$ with passing the symbol $\#$ to a prover. 
If the input is $\lambda$, then, in reaching the endmarker $\$$ in state $q_0$, $V$ enters the rejecting inner state $q_{rej}$. 
Clearly, $V$ rejects the input with certainty no matter how the prover behaves. 
In the opposite case where the input is nonempty, if $V$ scans $a$ for the first time in the initial inner state $q_0$, 
$V$ sends the symbol $a$ to a prover and then enters the inner state $q_1$.  When the honest prover modifies it back to $\#$, $V$ keeps the current inner state $q_1$ and the current communication symbol until $V$ 
reads $\$$. Finally, $V$ enters the accepting inner state $q_{acc}$. 
With the honest prover, $V$ correctly accepts the input with certainty. Hence, $(P,V)$ recognizes $L_a$ with certainty.

\begin{table}[ht]
\bs\begin{center}
\begin{tabular}{|ll|}\hline
$V_{\cent}\qubit{q_0}\qubit{\#}=\qubit{q_0}\qubit{\#}$ &  \\ 
$V_{a}\qubit{q_0}\qubit{\#}=\qubit{q_1}\qubit{a}$ & 
$V_{a}\qubit{q_1}\qubit{\#}=\qubit{q_1}\qubit{\#}$ \\
$V_{\$}\qubit{q_0}\qubit{b} =\qubit{q_{rej}}\qubit{b}$ & 
$V_{\$}\qubit{q_1}\qubit{\#} =\qubit{q_{acc}}\qubit{\#}$ \\  \hline
\end{tabular}
\caption{Transitions of $V$ for $L_a$ with $b\in \{a,\#\}$. 
The unitary operator $V_{\sigma}$ for each $\sigma\in\check{\Sigma}$ acts on the Hilbert space $\mathrm{span}\{\qubit{q,\gamma}\mid (q,\gamma)\in Q\times\Gamma\}$. The transition function $\delta$ of $V$ is then induced 
by letting $\delta(q,\sigma,\gamma,q',\gamma',1)=\langle q',\gamma'|V_\sigma|q,\gamma\rangle$ 
for every $q,q'\in Q$ and $\gamma,\gamma'\in\Gamma$.}\label{table:La}  
\end{center}
\end{table}

We next prove the remaining claim that $\{\lambda\}\not\in \qip(mo\mbox{-}1qfa)$. More generally, 
we claim that no finite language belongs to $\qip(mo\mbox{-}1qfa)$. 
This claim is a consequence of the following lemma, which gives a more 
general limit to the power of the QIP systems with mo-1qfa verifiers. 

\begin{figure}[ht]
\begin{center}
\centerline{\psfig{figure=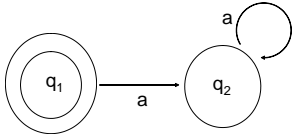,height=2cm}}
\caption{Transitions included in the minimal automaton for $L$}\label{fig:minimal}
\end{center}
\end{figure}

\begin{lemma}\label{mo-weak}
Let $L$ be a language over a nonempty alphabet $\Sigma$ and let $M$ be its minimal deterministic automaton. Assume that there 
exist an input symbol $a\in\Sigma$, an accepting inner state $q_1$, and a 
rejecting inner state $q_2$ satisfying: (1) if $M$ reads $a$ in the state $q_1$, then
$M$ enters the state $q_2$ and (2) if $M$ reads $a$ in the state $q_2$, then $M$ 
stays in the state $q_2$. Figure \ref{fig:minimal} illustrates these transitions. The language $L$ is then outside of $\qip(mo\mbox{-}1qfa)$.  
\end{lemma}

To prove Lemma \ref{mo-weak}, we use the following well-known result in \cite{BP02}.  

\begin{lemma}\label{BroPip99}{\rm \cite{BP02}}\hs{1}
Let $U$ be any unitary matrix and let $\epsilon$ be any positive real number. There exists a number $n\in\nat^{+}$ 
such that $\|(I-U^n)x\|^2<\epsilon$ for any vector $x$ with $\|x\|^2\leq 1$.
\end{lemma}

We give the proof of Lemma \ref{mo-weak}.

\begin{proofof}{Lemma \ref{mo-weak}}
From the characteristics of the minimal automaton, there exists an input string $y\in\Sigma^*$ such that $M$ enters $q_1$ 
after reading $y$ and enters $q_2$ after reading $ya^n$ for any positive integer $n$. Hereafter, we fix such a string $y$.
Assume toward a contradiction that $L$ belongs to $\mathrm{QIP}(mo\mbox{-}1qfa)$. Take a real number $\eta>0$, 
an honest prover $P$, and an mo-1qfa verifier $V$ satisfying the following: 
$(P,V)$ accepts $y$ with probability at least $1/2+\eta$ while, 
for any prover $P^*$ and any number $n\in\nat^{+}$, $(P^*,V)$ rejects $ya^n$ with probability $\geq 1/2+\eta$. 
Consider the following prover $P'$ that works on input $ya^n$: $P'$ first simulates $P$ on input $y$ 
while $V$ is reading $y$ and, whenever $V$ passes a symbol $s$ to $P'$, $P'$ returns the same $s$ to $V$. 

To lead to a contradiction, we utilize Lemma \ref{BroPip99}. Let $\qubit{\phi_y}$ be the superposition of configurations obtained after $V$ finishes reading $y$. 
Let $V_b$ be the unitary operator corresponding to the transition of $V$ while scanning symbol $b\in\check{\Sigma}$. 
By setting $\epsilon=\eta^2$, Lemma \ref{BroPip99} guarantees the existence of a positive integer $n$ such that $\|\qubit{\phi_y}-V_a^n\qubit{\phi_y}\|^2 <\eta^2$, which equals $\|\qubit{\phi_y}-V_a^n\qubit{\phi_y}\| <\eta$. For readability, we write $p_{acc}(y,P)$ for $p_{acc}(y,P,V)$.  Since $p_{acc}(y,P)$ and $p_{acc}(ya^n,P')$ 
are obtained respectively by measuring the final superpositions $V_{\$}\qubit{\phi_y}$ and $V_{\$}V_{a}^n\qubit{\phi_y}$, we conclude:
\[
|p_{acc}(y,P) - p_{acc}(ya^n,P')| \leq \|V_{\$}\qubit{\phi_y}- V_{\$}V_a^n\qubit{\phi_y}\| = \|\qubit{\phi_y}-V_a^n\qubit{\phi_y}\| <\eta,
\]
where the first inequality is a folklore (see, \eg \cite[Lemma 8]{Yam03}).
Since $p_{acc}(y,P)\geq 1/2+\eta$, it follows that $p_{acc}(ya^n,P')\geq (1/2+\eta) - \eta= 1/2$, 
which contradicts our assumption that, for any prover $P^*$, $p_{acc}(ya^n,P^*) \leq 1/2-\eta <1/2$. 
Therefore, $L\not\in\mathrm{QIP}(mo\mbox{-}1qfa)$. 
\end{proofof}

Earlier, Brodsky and Pippenger \cite{BP02} gave a group-theoretic characterization of $\mathrm{MO\mbox{-}1QFA}$. 
Such a characterization is not yet known for $\mathrm{QIP}(mo\mbox{-}1qfa)$.

\section{Is a Quantum Prover Stronger than a Classical Prover?}

Our prover can perform any operation that quantum physics allows. We want to restrict the power of a prover. If the 
prover is limited to wield only {\lq\lq}classical{\rq\rq} power, we may call 
such a prover {\lq\lq}classical.{\rq\rq} More precisely, a prover is called {\em classical} if 
the prover's move is dictated by a unitary operator whose entries are 
either $0$s or $1$s.  By contrast, we sometimes refer to any standard prover 
as a {\em quantum prover}. Remember that any classical prover is a quantum prover. 
Although any classical-prover QIP system seems to be directly simulated by a similar QIP system using a quantum prover, it is not yet known that this is truly the case in general because, 
intuitively, more powerful the prover becomes, more easily may the weak verifier be convinced as well as fooled. 
Hereafter, the restriction $\pair{c\mbox{-}prover}$ indicates that a prover behaves 
classically. In our qfa-verifier QIP systems, a classical prover may play an essentially different role from a quantum prover's. 

We consider the 1qfa-verifier case first. Similar to the quantum prover case, we can show that $\mathrm{1QFA}\subseteq\qip(1qfa,c\mbox{-}prover)$. By expanding this containment, we can show the following stronger containment, which makes a bridge between quantum provers and classical provers. 

\begin{proposition}\label{public-cprover}
$\qip(1qfa) \subseteq \qip(1qfa,c\mbox{-}prover)$.
\end{proposition}

\begin{proof}
It is easy to show in a way similar to Proposition \ref{regular-1qfa} that 
$\qip(1qfa,c\mbox{-}prover)$ contains all regular languages. 
Since $\qip(1qfa)=\mathrm{REG}$ by Theorem \ref{qfa-regular}, $\qip(1qfa,c\mbox{-}prover)$ therefore includes $\qip(1qfa)$. 
\end{proof}

Whether $\qip(1qfa,c\mbox{-}prover)$ coincides with $\qip(1qfa)$ is unclear due to the soundness condition of a QIP system. 

Next, we examine the 2qfa-verifier case.  
Unlike the 1qfa verifier case, any containment between $\qip(2qfa)$ and $\qip(2qfa,c\mbox{-}prover)$ is unknown. 
Nonetheless, we can easily show that 
$\qip(2qfa,poly\mbox{-}time,c\mbox{-}prover)$ contains $\mathrm{2QFA}(poly\mbox{-}time)$. 
The proper inclusion $\mathrm{REG} \subsetneqq \qip(2qfa,poly\mbox{-}time,c\mbox{-}prover)$ is a direct consequence of the result in \cite{KW97} that $\mathrm{REG}\subsetneqq\mathrm{2QFA}(poly\mbox{-}time)$.  
The following theorem greatly strengthens this separation.

\begin{theorem}\label{am-cprover}
1. $\am(2pfa)\subsetneqq \qip(2qfa,c\mbox{-}prover)$.

2. $\am(2pfa,poly\mbox{-}time)\subsetneqq \qip(2qfa,poly\mbox{-}time,c\mbox{-}prover)\nsubseteq \am(2pfa)$. 
\end{theorem}

\begin{proof}
In the proof of Lemma \ref{pal-sharp}, we have shown that $Pal_{\#}$ is in $\qip(2qfa,poly\mbox{-}time)$. Notice that the same proof works for classical provers. This places $Pal_{\#}$ in $\qip(2qfa,poly\mbox{-}time,c\mbox{-}prover)$. Hence, similar to Theorem \ref{am-polytime}, the separation between $\am(2pfa)$ and  $\qip(2qfa,poly\mbox{-}time,c\mbox{-}prover)$ naturally follows. 
This separation further leads to the inequality between $\am(2pfa)$ and $\qip(2qfa,c\mbox{-}prover)$ (also between $\am(2pfa,poly\mbox{-}time)$ and $\qip(2qfa,poly\mbox{-}time,c\mbox{-}prover)$).
Therefore, in this proof, it suffices to show that $\am(2pfa)\subseteq \qip(2qfa,c\mbox{-}prover)$. Since our proof works for any time-bounded case, we also obtain the remaining claim that $\am(2pfa,poly\mbox{-}time)\subseteq \qip(2qfa,poly\mbox{-}time,c\mbox{-}prover)$.

The important starting point is the fact that the complexity class $\am(2pfa)$ can be characterized 
by bounded-error finite automata with probabilistic and nondeterministic moves. 
Such an automaton is called  a {\em 2npfa} in \cite{CHPW98}. 
Let $L$ be any language in $\am(2pfa)$ over alphabet $\Sigma$. 
Take a finite automaton $M=(Q,\Sigma,\delta_{M})$ with nondeterministic states and probabilistic states that recognizes $L$ with error probability at most $\epsilon$, where $0\leq \epsilon < 1/2$. To simplify our proof, we make two inessential assumptions for $M$'s head move. Assume that (i) $M$'s head always moves either to the right or to the left and (ii) whenever $M$ tosses a fair coin, the head moves only to the right. Based on this $M$, we shall construct a QIP system $(P,V)$ for $L$.  

Let $x$ be any input of length $n$. 
The verifier $V$ carries out the following procedure, 
in which $V$ simulates $M$ step by step with $Q'=\{p,\hat{p}\mid p\in Q\}$ as the set of inner states 
and $\Gamma=(Q'\times\{\pm1\})\cup\{\#,\$\}$ as the communication alphabet, 
where $\hat{p}$ is a new inner state associated with $p$ and $\$$ is a new non-blank symbol. 
Consider any step at which $M$ tosses a fair coin in probabilistic state $p$ by the transition 
$\delta_M(p,\sigma)=\{(p_0,1),(p_1,1)\}$ for certain distinct states $p_0,p_1\in Q$. 
The verifier $V$ checks whether the communication cell is blank. 
If not, $V$ rejects $x$ at this simulation step; otherwise, 
$V$ makes the corresponding transition 
$V_\sigma\qubit{p}\qubit{\#} = \frac{1}{\sqrt{2}}(\qubit{p_0}\qubit{(p,1)}+ \qubit{p_1}\qubit{(p,1)})$. 
Here, $V_\sigma$ is the unitary operator defined by $\delta(p,\sigma,\gamma,q,\gamma',D(q))=
\langle q,\gamma'|V_\sigma|p,\gamma\rangle$ with the transition function $\delta$ of $V$ 
and $D$ is the function from $Q'$ to $\{0,\pm 1\}$. 
The verifier expects a prover to erase the symbol $p$ in the communication cell by overwriting it with the blank symbol $\#$. 
This erasure guarantees $V$'s move to be unitary. 

Next, consider any step at which $M$ makes a nondeterministic choice in state $p$ 
by the transition $\delta_M(p,\sigma)=\{(p_0,d_0),(p_1,d_1),\ldots,(p_m,d_m)\}$, where $m\in\nat$. 
Notice that a deterministic move is treated as a special case of a nondeterministic move. 
In this case, $V$ takes two steps to simulate $M$'s move. 
The verifier $V$ enters a rejecting inner state immediately unless the communication cell contains the blank symbol. 
Now, assume that the communication cell is blank. 
Without moving its head, $V$ first sends the designated symbol $\$$ to a prover, 
requesting a pair $(p',d')$ in $Q\times\{\pm 1\}$ to return. 
This is done by the transition $V_\sigma\qubit{p}\qubit{\#}=\qubit{\hat{p}}\qubit{\$}$. 
The verifier forces a prover to return a valid nondeterministic choice (\ie $(p',d')\in\delta_{M}(p,\sigma)$) 
by entering a rejecting inner state if the prover writes any other symbol. 
Once $V$ receives a valid pair $(p',d')$, 
$V$ makes the transition $V_\sigma\qubit{\hat{p}}\qubit{(p',d')} = \qubit{p'}\qubit{(\hat{p},d')}$ 
and expects a prover to erase the communication symbol $(\hat{p},d')$.

The honest prover $P$ must blank the communication cell at the end of each simulation step of $V$ 
and return a {\lq\lq}correct{\rq\rq} nondeterministic choice on request of the verifier $V$.
If $x\in L$, there are a series of nondeterministic choices along which $M$ accepts $x$ with probability at least $1-\epsilon$. With the help of the honest prover $P$, $V$ can successfully simulate $M$  with the same error probability. Consider the case where $x\not\in L$, on the contrary. In this case, no matter how nondeterministic choices are made, $M$ rejects $x$ with probability at least $1-\epsilon$. Take a dishonest classical prover $P^*$ that maximizes the acceptance probability of $V$ on $x$. 
This prover $P^*$ must clear out the communication cell whenever $V$ asks him to do so since, 
otherwise, $V$ immediately rejects $x$. Since $P^*$ is classical, all the computation paths of $V$ 
have nonnegative amplitudes which cause only constructive interference. 
This indicates that $P^*$ cannot annihilate any existing computation path of $V$. 
On request for a nondeterministic choice, $P^*$ must return any one of valid nondeterministic choices. 
With a series of nondeterministic choices of $P^*$, if $V$ rejects $x$ with probability less than $1-\epsilon$, 
then our simulation implies that $M$ rejects $x$ with probability less than $1-\epsilon$. 
This is a contradiction against our assumption. Hence, $V$ rejects $x$ with probability at least $1-\epsilon$. 
Therefore, $(P,V)$ is a $(1-\epsilon,1-\epsilon)$-QIP system for $L$.
\end{proof}

In the above proof, we cannot replace a classical prover by a quantum prover. 
The major reason is that a quantum prover may (i) return a superposition of two nondeterministic choices instead of choosing one of the two choices and 
(ii) use negative amplitudes to make the verifier's quantum simulation destructive.

In the end of this section, we present a QIP protocol with a classical prover for the non-regular language $Center$, 
which is known to be in $\am(2pfa)$ but not in $\am(2pfa,poly\mbox{-}time)$ \cite{DS92}. 
In our QIP system, a prover signals the location of the center bit of an input 
and then a verifier tests the correctness of the location by employing the quantum Fourier transformation (QFT, in short) 
in a fashion similar to \cite{KW97}. 

\begin{lemma}\label{qip-vs-am}
For any $\epsilon\in(0,1)$, $Center\in \qip_{1,1-\epsilon}(2qfa,poly\mbox{-}time,c\mbox{-}prover)$. 
\end{lemma}

\begin{proof}
Let $\epsilon$ be any error bound in the real interval $(0,1)$ and set $N=\ceilings{1/\epsilon}$. 
We give a QIP protocol witnessing the membership of $Center$ to $\qip_{1,1-\epsilon}(2qfa,poly\mbox{-}time,c\mbox{-}prover)$.
Let $\Sigma=\{0,1\}$ be our input alphabet and let $\Gamma=\{\#,1\}$ be our communication alphabet. 
Our QIP protocol comprises four phases. 
Let $x$ be an arbitrary input.
In the first phase, the verifier checks whether $|x|$ is odd by moving the head toward the right endmarker $\$$ 
together with switching two inner states $q_0$ and $q_1$. 
To make deterministic moves, the verifier forces a prover to return only the blank symbol $\#$. 
When $|x|$ is odd, the verifier enters the state $q_3$ after stepping back to $\cent$. 
Hereafter, we consider only the case where input $x$ has an odd length.

In the second phase, $V$ moves its head rightward by passing the communication symbol $\#$ to a prover 
until $V$ receives $1$ from the prover. 
Receiving $1$ from the prover, $V$ rejects $x$ unless scanning $1$ in the input tape. 
Otherwise, the third phase starts. During the third and fourth phases, 
whenever the prover changes the communication symbol $1$ to $\#$, $V$ immediately rejects the input. Assume that the head is now scanning $1$. In the third phase, the computation splits into $N$ parallel branches (the first split) 
generating the $N$ distinct inner states $r_{1,0},r_{2,0},\ldots,r_{N,0}$ with equal amplitudes $1/\sqrt{N}$. 
The head then moves deterministically toward the right endmarker $\$$ in the following manner: 
along the $j$th path ($1\leq j\leq N$) associated with the inner state $r_{j,0}$, 
the head idles for $2(N-j)$ steps in each tape cell before moving to the next one. 
When the head reaches $\$$, it steps back two cells and starts the fourth phase. 
During the fourth phase, the head along the $j$th path keeps moving leftward by idling in each cell for $j$ steps 
until the head reaches  $\cent$.
At the left endmarker, the computation splits again into $N$ parallel branches by the QFT  
(the second split), yielding either the accepting inner state $t_N$ or one of the rejecting inner states 
$\{t_j\mid 1\leq j< N\}$. 

The formal description of the transitions of $V$ is given in Table \ref{table:center}. {}From this table, it is not difficult to check that the verifier is well-formed (\ie $U^x_{\delta}$ is unitary for every $x\in\Sigma^*$). 
The honest prover $P$ should return $1$ exactly at the time when the verifier scans the center bit of an input 
and at the time when the verifier sends $\#$ during the third and fourth phases.  
At any other step, $P$ should perform the identity operation. 

\begin{table}[ht]
\bs\begin{center}
{\small
\begin{tabular}{|ll|}\hline
$V_{\cent}\qubit{q_0}\qubit{\#} = \qubit{q_0}\qubit{\#}$  
& $V_{\$}\qubit{q_0}\qubit{\#} =\qubit{q_{rej,0}}\qubit{\#}$  \\ 

$V_{\cent}\qubit{q_2}\qubit{1}=\qubit{q_{rej,0}}\qubit{\#}$ 
& $V_{\$}\qubit{q_1}\qubit{\#} =\qubit{q_2}\qubit{\#}$ \\

$V_{\cent}\qubit{s_{j,0}}\qubit{1} = \frac{1}{\sqrt{N}}\sum_{l=1}^N \mathrm{exp}(2\pi \imath jl/{N})
\qubit{t_{l}}\qubit{\#}$ ($1\leq j\leq N$) 
& $V_{\$}\qubit{q_0}\qubit{1} =\qubit{q_{rej,1}}\qubit{\#}$\\

$V_{\cent}\qubit{q_2}\qubit{\#}=\qubit{q_3}\qubit{\#}$ 
& $V_{\$}\qubit{q_1}\qubit{1} =\qubit{q_{rej,1}}\qubit{1}$\\

&  $V_{\$}\qubit{r_{j,0}}\qubit{1} = \qubit{s'_{j,0}}\qubit{1}$  ($1\leq j\leq N$) \\ 

$V_b\qubit{q_0}\qubit{\#}=\qubit{q_1}\qubit{\#}$ 
& $V_b\qubit{q_1}\qubit{\#}=\qubit{q_0}\qubit{\#}$ \\

$V_b\qubit{q_0}\qubit{1} = \qubit{q_{rej,0}}\qubit{\#}$ 
& $V_b\qubit{q_1}\qubit{1} = \qubit{q_{rej,0}}\qubit{1}$ \\

$V_b\qubit{q_2}\qubit{\#} = \qubit{q_2}\qubit{\#}$ & $V_b\qubit{q_2}\qubit{1}=\qubit{q_{rej,1}}\qubit{1}$ 
 \\

$V_b\qubit{q_3}\qubit{\#}=\qubit{q_3}\qubit{\#}$ & \\

$V_1\qubit{q_3}\qubit{1} = \frac{1}{\sqrt{N}}\sum_{j=1}^N \qubit{r_{j,0}}\qubit{\#}$ 
& $V_0\qubit{q_3}\qubit{1} = \qubit{q_{rej,-1}}\qubit{\#}$ \\ 

$V_b\qubit{r_{j,0}}\qubit{1} = \qubit{r'_{j,N-j}}\qubit{1}$ ($1\leq j\leq N-1$) 
& $V_b\qubit{r_{j,0}}\qubit{\#}=\qubit{q_{rej,j}}\qubit{1}$ ($1\leq j\leq N-1$) \\

$V_b\qubit{r_{j,k}}\qubit{1} = \qubit{r'_{j,k}}\qubit{1}$ ($1\leq k\leq N-j$, $1\leq j\leq N-1$) & \\

$V_b\qubit{r'_{j,k}}\qubit{1} = \qubit{r_{j,k-1}}\qubit{1}$ ($2\leq k\leq N-j$, $1\leq j\leq N-1$)  & \\

$V_b\qubit{r'_{j,1}}\qubit{1} = \qubit{r_{j,0}}\qubit{1}$, ($1\leq j\leq N$)  
& $V_b\qubit{r_{N,0}}\qubit{1} = \qubit{r_{N,0}}\qubit{1}$ \\

$V_b\qubit{s'_{j,0}}\qubit{1} = \qubit{s_{j,0}}\qubit{\#}$ ($1\leq j\leq N$) 
& $V_b\qubit{s_{j,0}}\qubit{1} = \qubit{s_{j,j}}\qubit{1}$ ($1\leq j\leq N$) \\ 

$V_b\qubit{s_{j,k}}\qubit{1}=\qubit{s_{j,k-1}}\qubit{1}$ ($2\leq k\leq j$, $1\leq j\leq N$)  & \\

$V_b\qubit{s_{j,1}}\qubit{1}=\qubit{s_{j,0}}\qubit{1}$ ($1\leq j\leq N$)  
& $V_b\qubit{s_{j,0}}\qubit{\#}=\qubit{q_{rej,N+j}}\qubit{\#}$ ($1\leq j\leq N$)\\

& \\

$D(q_0)=D(q_1)=D(q_3)=1$, $D(q_2)=-1$ & $D(r_{j,0})=1$ ($1\leq j\leq N$) \\ 
$D(r_{j,k})=D(r'_{j,k})=0$ ($1\leq j\leq N-1$, $k\neq 0$)  
& $D(s_{j,0})=D(s'_{j,0})=-1$ ($1\leq j\leq N$) \\
 $D(s_{j,k})=0$ ($1\leq j\leq N$, $k\neq 0$)  & $D(t_j)=0$ ($1\leq j\leq N$) \\ 
\hline
\end{tabular}
}
\caption{Transitions of $V$ for $Center$ with $b\in\{0,1\}$. 
In this table, $t_N$ is the only accepting inner state 
while $q_{rej,j}$ ($-1\leq j\leq 2N-1$) and $t_l$ ($1\leq l<N$) are rejecting inner states. 
The table, however, excludes obvious transitions to rejecting inner states when a prover changes the communication symbol $1$ to $\#$ during the third and fourth phases. The transition function $\delta$ is induced from $V$ as $\delta(q,\sigma,\gamma,q',\gamma',d)
=\langle q',\gamma'|V_\sigma|q,\gamma\rangle$ if $D(q')=d$ and $0$ otherwise.}\label{table:center}
\end{center}
\end{table}
 
The following is the proof of the completeness and soundness of the QIP system $(P,V)$ for $Center$. 
First, consider a positive instance $x$, which is of the form $y1z$ for certain strings $y$ and $z$ of the same length, 
say $n$. 
Since the honest prover $P$ signals when the verifier reads the center bit of $x$, 
the first split occurs exactly after $n$ steps of $V$ from the start of the second phase. 
Along the $j$th path ($1\leq j\leq N$) chosen at the first split, 
$V$ idles for $2n(N-j)$ steps while reading $y$ and also idles for $(|x|-1)j$ steps while reading the whole input 
except for its rightmost symbol. 
Overall, the idling time elapses for the duration of $2n(N-j)+2nj=2nN$, which is independent of $j$. 
Hence, all the $N^2$ paths created at the two splits have the same length. 
The QFT then converges them to the verifier's visible accepting configuration $\qubit{t_N}\qubit{\#}$. 
Therefore, $V$ accepts $x$ with probability 1. 

On the contrary, suppose that $x=y0z$, where $|y|=|z|=n$. Consider the second, third and fourth phases.
To minimize the rejection probability, a dishonest prover $P^*$ should send the symbol $1$ at the moment when $V$ scans $1$ in the input tape in the second phase and then maintain $1$ after the first split because, otherwise, 
$V$ immediately rejects $x$ and no classical prover passes both $1$ and $\#$ in a form of superposition.  
Now assume that the $e$th symbol of $x$ is 1 and $P^*$ sends $1$ during the $e$th interaction, where $1\leq e\leq 2n+1$. 
Note that $e\neq n+1$ because the center bit of $x$ is $0$. For any $j\in[1,N]_{\integer}$, let $p_{j}$ be the computation path 
following the $j$th branch generated at the first split.  
Along this path $p_{j}$ toward the left endmarker $\cent$, the idling time totals $2(|x|-e)(N-j)+2nj=2(n+1-e)(N-j)+2nN$. 
For any distinct values $j$ and $j'$, the two paths $p_{j}$ and $p_{j'}$ have different lengths. 
For each of such paths, the QFT further generates $N$ parallel paths; however, 
only one of them reach $\qubit{t_N}\qubit{\#}$. Hence, the probability of $V$ reaching such an acceptance configuration is no more than $1/N^2$. 
Since there are $N$ paths $\{p_j\}_{1\leq j\leq N}$, the overall acceptance probability is at most $N\times(1/N^2)=1/N$. 
It is easy to see that $V$ rejects $x$ with probability $\geq 1- 1/N \geq 1-\epsilon$.    
\end{proof}

\section{What If a Verifier Reveals His Private Information?}

The strength of a prover's strategy hinges on the amount of the information 
that a verifier reveals. For instance, when a verifier makes only one-way communication (as in the proof of Lemma \ref{pal-sharp}), no prover gains more than the information on the number of the verifier's moves. The prover therefore knows little of the verifier's configurations. In Babai's {\lq\lq}public{\rq\rq} IP systems by contrast, a verifier completely reveals his configurations. The notion of {\lq\lq}public coins{\rq\rq} forces the verifier to pass {\em only} his choice of next moves, which allows the prover to reconstruct the verifier's computation. In this section, we consider a 
straightforward analogy of public IP systems in the quantum setting and call our QIP system {\em public} for convenience. Formally, we introduce a {\em public QIP system} as follows. 

\begin{definition}
A qfa-verifier QIP system $(P,V)$ is called {\em public} if 
the verifier $V$ writes his choice of non-halting inner state and head direction in 
the communication cell at every step; that is, the verifier's transition 
function $\delta$ satisfies that, for any $x$, $q$, $k$, and $\gamma$, $U_\delta^x\qubit{q,k,\gamma}= 
\sum_{q',\xi,d} \delta(q,x_{(k)},\gamma,q',\xi,d)\qubit{q',k+d\ (\mbox{mod }n+2),\xi}$, 
where $\xi=(q',d)$ whenever $q'$ is a non-halting inner state.
\end{definition}

In particular, when the verifier $V$ is a 1qfa, we can omit the head-direction information $d$ from the communication symbol $\xi=(q',d)$ in the above definition since $V$ always moves its head to the right. To emphasize the public QIP system, we use the restriction 
$\pair{public}$. 

Let us begin our study on the complexity 
class $\qip(1qfa,public)$. 

\begin{proposition}\label{public-1qfa}
$\qip_{1,1}(1qfa,public) \nsubseteq \mathrm{1QFA}$. 
\end{proposition}

Recall the language $Zero$. 
Proposition \ref{public-1qfa} is obtained by proving that $Zero$ belongs to $\qip(1qfa,public)$ 
since $Zero$ resides outside of $\mathrm{1QFA}$ \cite{KW97}. 
The following proof exploits the prover's ability to inform the location of the rightmost bit $0$ of an instance in $Zero$.    

\begin{proofof}{Proposition \ref{public-1qfa}}
We want to show that $Zero$ has an error-free public QIP system $(P,V)$ with a 1qfa verifier. 
Since no 1qfa recognizes the language $Zero$ \cite{KW97}, we therefore obtain the proposition. 
To describe the desired protocol $(P,V)$, let $\Sigma=\{0,1\}$ be its input alphabet and let  
$Q_{non}=\{q_0,q_1\}$, $Q_{acc}=\{q_{acc,0},q_{acc,1},q_{acc,-1}\}$ and $Q_{rej}=\{q_{rej,0},q_{rej,1},q_{rej,-1}\}$ 
be respectively the sets of all non-halting inner states, accepting inner states and rejecting inner states of $V$. 

As mentioned before, we abbreviate communication symbol $(q,1)$ for $q\in Q$ as $q$ since $V$'s head direction is always $+1$. Our communication alphabet $\Gamma$ is thus $\{\#,q_0,q_1\}$. 
The protocol of $V$ is described in the following. 
Let $x=yb$ be any input string, where $b\in\{0,1\}$. 
The verifier $V$ stays in the initial state $q_0$ by sending the communication symbol $q_0$ to a prover until the prover returns $\#$. Whenever $V$ receives $\#$, he immediately rejects $x$ if its current scanning symbol is different from $0$.  On the contrary, if $V$ is scanning $0$, then he waits for the next tape symbol. If the next symbol is $\$$, then he accepts $x$; otherwise, he rejects $x$. See Table \ref{table:public-Lzero} for the formal description of $V$'s transitions.  Our honest prover $P$ does not alter the communication cell until $V$ reaches the end of $\cent y$ and he must return $\#$ exactly when $V$ reads the rightmost symbol of $\cent y$. 

\begin{table}[ht]
\bs\begin{center}
\begin{tabular}{|lll|}\hline
$V_{\cent}\qubit{q_0}\qubit{\#}=\qubit{q_0}\qubit{q_0}$ 
& $V_{1}\qubit{q_0}\qubit{q_0}=\qubit{q_0}\qubit{q_0}$ & 
$V_{0}\qubit{q_0}\qubit{q_0}=\qubit{q_0}\qubit{q_0}$ \\  
$V_{\$}\qubit{q_0}\qubit{q_i} =\qubit{q_{rej,i}}\qubit{\#}$ & 
$V_{1}\qubit{q_0}\qubit{\#}=\qubit{q_{rej,-1}}\qubit{\#}$ & 
$V_{0}\qubit{q_0}\qubit{\#}=\qubit{q_{1}}\qubit{q_{1}}$ \\ 
$V_{\$}\qubit{q_{1}}\qubit{q_i} =\qubit{q_{acc,i}}\qubit{\#}$ & 
$V_{1}\qubit{q_{1}}\qubit{q_i} =\qubit{q_{rej,i}}\qubit{q_0}$ & 
$V_{0}\qubit{q_{1}}\qubit{q_i} =\qubit{q_{rej,i}}\qubit{q_0}$ \\
& $V_{1}\qubit{q_0}\qubit{q_1}=\qubit{q_{rej,1}}\qubit{\#}$ & 
$V_{0}\qubit{q_0}\qubit{q_1}=\qubit{q_{rej,1}}\qubit{\#}$ \\  \hline
\end{tabular}
\caption{Transitions of $V$ for $Zero$ with $i\in \{0,\pm 1\}$. The symbol $q_{-1}$ denotes $\#$.}\label{table:public-Lzero}  
\end{center}
\end{table}

It still remains to prove that $(P,V)$ recognizes $Zero$ with certainty. Consider the case where our input $x$ is of the form $y0$ for a certain string $y$. Since $x\in Zero$, the honest prover $P$ returns $\#$ exactly when $V$ reads the rightmost symbol of $\cent y$. This information helps $V$ locate the end of $y$.  
Now, $V$ confirms that the current scanning symbol is $0$ and then enters an accepting inner state with probability $1$ after it encounters the right endmarker. On the contrary, assume that $x=y1$. 
Clearly, the best adversary $P^*$ needs to return either $q_0$ or $\#$ (or their superposition). 
If $P^*$ keeps returning $q_0$, then $V$ eventually rejects $x$ and increases the rejection probability. 
Since $V$'s computation is deterministic, this only weakens the strategy of $P^*$. 
To make the best of the adversary's strategy, $P^*$ must return the communication symbol $\#$ before $V$ reaches $\$$. 
Nonetheless, although $P^*$ returns it, $V$ is designed to lead to a rejecting inner state. 
Therefore, the QIP system $(P,V)$ recognizes $Zero$ with certainty.   
\end{proofof}

A {\em 1-way reversible finite automaton} ({\em 1rfa}, in short) is a 1qfa whose transition amplitudes are either 0 or 1. 
Let $\mathrm{1RFA}$ denote the collection of all languages recognized by certain 1rfa's.
As Ambainis and Freivalds \cite{AF98} showed, $\mathrm{1RFA}$ is characterized as the collection of all languages 
that can be recognized by 1qfa's with success probability $\geq 7/9+\epsilon$ for certain numbers $\epsilon>0$.

\begin{proposition}
$\mathrm{1RFA}\subsetneqq \qip_{1,1}(1qfa,public)$. 
\end{proposition}

\begin{proof}
We first show that $\mathrm{1RFA}\subseteq \qip_{1,1}(1qfa,public)$. 
Take an arbitrary set $L$ recognized by a 1rfa $M=(Q,\Sigma,q_0,Q_{acc},Q_{rej},\delta_M)$. 
Without loss of generality, we can assume that, in the transition of $M$, 
the initial state $q_0$ appears only when $M$ starts its computation. 

\begin{table}[ht]
\bs\begin{center}
\begin{tabular}{|l|}\hline
$V_{\cent}\qubit{q_0}\qubit{\#}=\qubit{q}\qubit{q}$ if $\delta_{M}(q_0,\cent)=q$ \\
 
$V_{b}\qubit{p}\qubit{p}=\qubit{q}\qubit{q}$ if $\delta_{M}(p,b)=q\in Q_{non}$ \\
 
$V_{b}\qubit{p}\qubit{p}=\qubit{q}\qubit{\#}$ if $\delta_{M}(p,b)=q\in Q_{acc}\cup Q_{rej}$ and $b\neq\cent$ \\  
\hline
\end{tabular}
\caption{Transitions of $V$ for $L$ with $b\in\Sigma$ and $p,q\in Q$}\label{table:public-cprover}  
\end{center}
\end{table}

The protocol of $V$ is given as follows. 
Assume that $V$ is in inner state $p$ scanning symbol $b$. 
Whenever $M$ changes its inner state from $p$ to $q$ while scanning $b$, 
$V$ does so by sending the communication symbol $p$ to a prover if $q$ is a non-halting inner state. As soon as $V$ finds that the communication symbol has been altered by the prover, $V$ immediately rejects the input. Table \ref{table:public-cprover} gives the list of $V$'s unitary operators induced from $M$'s transition function $\delta_M$. The honest prover $P$ is the one who does not alter any communication symbol. On any input $x$,  the QIP system $(P,V)$ clearly accepts $x$ with certainty if $x\in L$. Consider the opposite case where $x\not\in L$. 
It is easy to see that the best strategy for a dishonest classical prover $P^*$ is to keep any communication symbol unchanged 
because any alteration of a communication symbol causes $V$ to reject $x$ immediately. 
Even with such a prover $P^*$, $V$ rejects $x$ with certainty. 
Therefore, $(P,V)$ recognizes $L$ with certainty. 
Since $L$ is arbitrary, we obtain the desired inclusion $\mathrm{1RFA}\subseteq\qip_{1,1}(1qfa,public)$. 
Finally, the separation between $\mathrm{1RFA}$ and $\qip_{1,1}(1qfa,public)$ comes from Proposition \ref{public-1qfa}. 
This completes the proof. 
\end{proof}

We further examine public QIP systems with 2qfa verifiers. 
Similar to Theorem \ref{am-cprover}(2), we can give the following separation.   

\begin{theorem}\label{public-vs-am}
\begin{enumerate}\vs{-2}
\item $\qip(2qfa,public,poly\mbox{-}time) \nsubseteq\am(2pfa,poly\mbox{-}time)$.
\vs{-2}
\item $\qip(2qfa,public,poly\mbox{-}time,c\mbox{-}prover) \nsubseteq\am(2pfa,poly\mbox{-}time)$.
\end{enumerate}
\end{theorem}

\sloppy
A language that separates the public QIP systems from $\am(2pfa,poly\mbox{-}time)$ is $Upal$. Since  $Upal$ resides outside of $\am(2pfa,poly\mbox{-}time)$ \cite{DS92} and $Upal$ belongs to $\mathrm{2QFA}(poly\mbox{-}time)$ \cite{KW97}, the separation $\mathrm{2QFA}(poly\mbox{-}time)\nsubseteq 
\am(2qfa,poly\mbox{-}time)$ follows immediately. This separation, however, does not directly imply Theorem \ref{public-vs-am} because it is not clear whether $\mathrm{2QFA}(poly\mbox{-}time)$ 
is included in $\qip(2qfa,public,poly\mbox{-}time)$ or in $\qip(2qfa,public,poly\mbox{-}time,c\mbox{-}prover)$. Therefore, we still need to prove in Lemma \ref{upal-public} that $Upal$ is indeed in both $\qip(2qfa,public,poly\mbox{-}time)$ and $\qip(2qfa,public,poly\mbox{-}time,c\mbox{-}prover)$. Our public QIP system for $Upal$, nevertheless, is essentially a slight modification of the 2qfa given in \cite{KW97} for $Upal$.

\begin{lemma}\label{upal-public}
\sloppy
For any constant $\epsilon\in(0,1]$,
$Upal\in \qip_{1,1-\epsilon}(2qfa,public,poly\mbox{-}time) \cap \qip_{1,1-\epsilon}(2qfa,public,poly\mbox{-}time,c\mbox{-}prover)$.
\end{lemma}

\begin{proof}
We show that $Upal$ belongs to $\qip_{1,1-\epsilon}(2qfa,public,poly\mbox{-}time)$ since the proof that $Upal$ belongs to $\qip_{1,1-\epsilon}(2qfa,public,poly\mbox{-}time,c\mbox{-}prover)$ is similar.   
Let $N=\ceilings{1/\epsilon}$. We define our public QIP system $(P,V)$ as follows. The verifier $V$ acts as follows. In the first phase, it determines whether an input $x$ is of the form $0^m1^n$. The rest of the verifier's algorithm is similar in essence to the one given in the proof of Lemma \ref{qip-vs-am}. In the second phase, $V$ generates $N$ branches with amplitude $1/\sqrt{N}$ by entering $N$ different inner states, say $r_{1},r_{2},\ldots,r_{N}$. 
In the third phase, along the $j$th branch starting with $r_j$ ($j\in[1,N]_\integer$), the head idles for $N-j$ steps at each tape cell containing $0$ and idles for $j$ steps at each cell containing $1$ until the head finishes reading $1$s. In the fourth phase, $V$ applies the QFT to collapse all the paths to a single accepting inner state if $m=n$. Otherwise, all the paths do not interfere with each other since the head reaches the right endmarker at different times along different branches. During the first and second phases, $V$ publicly reveals the information $(q',d')$ on his next move and then checks whether the prover rewrites it with a different symbol. 
To constrain the prover's strategy, $V$ immediately enters a rejecting inner state 
if the prover alters the content of the communication cell. 
The honest prover $P$ always applies the identity operation at every step. 

We show the completeness and soundness for our QIP system $(P,V)$. This is done in a fashion similar to the proof of Lemma \ref{qip-vs-am}. With the honest prover for any input $x\in Upal$, $(P,V)$ obviously accepts $x$ with probability $1$. Assume that $x=0^m1^n$ with $m\neq n$. 
Consider a dishonest prover $P^*$ who maximizes the acceptance probability of $V$ on $x$. 
Against $V$'s rejection criteria, the prover $P^*$ cannot change the content of the communication cell at any step. Since the head arrives at the endmarker $\$$ at different moments, no two branches apply the QFT simultaneously. This makes it impossible for $P^*$ to force two or more branches to interfere. Along each branch, the probability that $V$ enters an accepting inner state is at most $1/N^2$. Therefore, $(P^*,V)$ rejects $x$ with probability bounded below by $1-N\cdot(1/N^2)$, which is at least $1-\epsilon$. 
\end{proof}

As noted in the proof of Theorem \ref{am-cprover}, the classical public IP systems with 2pfa verifiers can be characterized by alternating automata 
that make nondeterministic moves and probabilistic moves. A natural question is whether our public QIP systems have a similar characterization in terms of a certain variation of qfa's. Moreover, Condon \etalc~\cite{CHPW98} proved that any language in $\am(2pfa,poly\mbox{-}time)$ has polylogarithmic 1-tiling complexity. What is the 1-tiling complexity of languages in 
$\qip(2qfa,public,poly\mbox{-}time)$? 

\section{How Many Interactions are Necessary?}
\label{sec:interaction}

In the previous sections, we have shown that quantum 
interactions between 
a prover and a qfa verifier notably enhance the qfa's 
ability to recognize certain types of languages. Since 
our basic model of QIP systems forces a verifier to communicate 
with a prover at every move, it is natural to ask whether 
such interactions 
are truly necessary. Throughout this section, we carefully 
examine the number of 
interactions between a prover and a verifier in a QIP system. 
To study such a number, 
we need to modify our basic systems so that a prover should 
alter a communication symbol in the communication cell exactly when 
the verifier asks the prover to do so. For such a modification, we first 
look into the IP systems of Dwork and Stockmeyer \cite{DS92}. In their 
system, a verifier is allowed to do computation silently at 
any chosen time with no communication 
with a prover. The verifier interacts with the prover 
only when the help of the prover is needed. We interpret the verifier's 
silent mode as follows: if the verifier $V$ does not wish to communicate 
with the prover, he writes a special communication symbol  
in the communication cell to signal the prover that he needs no 
help from the prover. Simply, we use the blank symbol $\#$ to condition that 
the prover is prohibited to tailor the content of the communication cell. 

We formally introduce a new QIP system, in which no malicious 
prover $P$ is permitted to cheat a verifier by tampering with 
the symbol $\#$ 
willfully. To describe a {\lq\lq}valid{\rq\rq} prover $P$ 
independent of the choice of a verifier, we require the prover's strategy 
$P_x=\{U_{P,i}^x\}_{i\in\nat^+}$ on input $x$, acting on the 
prover's visible configuration space $\MM\otimes\PP$, to satisfy 
the following condition. For each $i\in\nat$, let $S_0=\{\#^{\infty}\}$ 
and let $S_{i}$ be the collection of all $y\in \Delta^{\infty}_{fin}$ 
such that, for a certain element $z\in S_{i-1}$ and certain 
communication symbols $\sigma,\tau\in\Gamma^{*}$, 
the superposition $U^{x}_{P,i}\qubit{\sigma}\qubit{z}$ contains 
the configuration $\qubit{\tau}\qubit{y}$ of non-zero amplitude. 
Note that these $S_i$'s are all finite. For every $i\in\nat^{+}$ 
and every $y\in S_{i-1}$, we require the existence of a pure quantum 
state $\qubit{\psi_{x,y,i}}$ in the Hilbert space spanned by 
$\{\qubit{z}\mid z\in \Delta^{\infty}_{fin}\}$ for which 
$U_{P,i}^x\qubit{\#}\qubit{y}=\qubit{\#}\qubit{\psi_{x,y,i}}$. 
A prover who meets this condition is briefly referred to as {\em 
committed}. A trivial example of such a committed prover is the 
prover $P_{I}$, who always applies the identity operation. A 
committed prover lets the verifier safely make a number of moves 
without any {\lq\lq}direct{\rq\rq} interaction with him. Observe 
that this new model with committed provers is in essence close to 
the circuit-based QIP model 
discussed in Section \ref{sec:circuit}. We name our new model 
an {\em 
interaction-bounded QIP system} and use the new notation 
$\qip^{\#}(1qfa)$ for the class of all languages recognized 
with bounded error by such interaction-bounded QIP systems 
with 1qfa verifiers. Since $\qip^{\#}(1qfa)$ naturally contains 
$\qip(1qfa)$, our interaction-bounded QIP systems can also 
recognize the 
regular languages. This simple fact will be used later.

\begin{lemma}\label{qipregular}
$\mathrm{REG}\subseteq \qip^\#(1qfa)$. 
\end{lemma}

Next, we need to clarify the meaning of the number of interactions. 
Consider any non-halting global configuration in which
$V$ on input $x$ communicates with a prover (\ie writes a non-blank symbol in the communication cell). 
For convenience, we call such a global configuration a {\em query configuration} and, at a query configuration, $V$ is said to {\em query} a word to a prover.
The {\em number of interactions} in a given computation means the maximum number, over all computation paths $\gamma$, of all the query configurations of non-zero amplitudes along its computation path $\gamma$. Let $L$ be any language and let $(P,V)$ be any interaction-bounded QIP system recognizing $L$. We say that the QIP protocol {\em $(P,V)$ makes $i$ interactions} on input $x$ if $i$ equals the number of interactions during the computation of $(P,V)$ on $x$. Furthermore, we call $(P,V)$ {\em $k$-interaction bounded} if, for every $x$, if $x\in L$ then $(P,V)$ makes at most $k$ interactions on input $x$\footnote{Instead, we may possibly consider a stronger condition like: for every $x$ and every committed prover $P^*$, $(P^*,V)$ makes at most $k$ interactions.} and otherwise, for every committed prover $P^*$, $(P^*,V)$ makes at most $k$ interactions on input $x$. At last, let $\qip^{\#}_{k}(1qfa)$ denote the class 
of all languages recognized with bounded error by $k$-interaction bounded QIP systems with 1qfa verifiers. Obviously, $\mathrm{1QFA}\subseteq \qip^{\#}_{k}(1qfa)\subseteq \qip^{\#}_{k+1}(1qfa)\subseteq \qip^{\#}(1qfa)$ for any number $k\in\nat$. In particular, $\qip^{\#}_{0}(1qfa) = \mathrm{1QFA}$.

As the main theorem of this section, we show in Theorem \ref{query-once} that (i) $1$-iteration helps a 
verifier but (ii) $1$-iteration does not achieve the full power of $\qip^{\#}(1qfa)$. 

\begin{theorem}\label{query-once}
$\qip^{\#}_0(1qfa) \subsetneqq\qip^{\#}_1(1qfa) 
\subsetneqq\qip^{\#}(1qfa)$. 
\end{theorem}

Theorem \ref{query-once} is a direct consequence of Lemma \ref{odd-1qfa} and Proposition \ref{zero-1qfa}. For the first inequality of Theorem \ref{query-once}, we use the language $Odd$ defined as 
the set of all binary strings of the form $0^m1z$, where $m\in\nat$, 
$z\in\{0,1\}^*$, and $z$ contains an odd number of $0$s. Since $Odd\not\in \mathrm{1QFA}$ \cite{AKV01}, it is enough for us to show in Lemma \ref{odd-1qfa} that $Odd$ belongs to $\qip^{\#}_1(1qfa)$. 
For the second inequality, we shall demonstrate in Proposition \ref{zero-1qfa} that  $\qip^{\#}_1(1qfa)$ does not include the regular language $Zero$. Since $\mathrm{REG}\subseteq\qip^{\#}(1qfa)$ by Lemma \ref{qipregular}, $Zero$ belongs to $\qip^{\#}(1qfa)$ and we therefore obtain the desired separation. 

The rest of this section is devoted to prove Lemma \ref{odd-1qfa} and Proposition \ref{zero-1qfa}. As the first step, we prove Lemma \ref{odd-1qfa}.

\begin{lemma}\label{odd-1qfa}
$Odd\in\qip^{\#}_{1}(1qfa)$.
\end{lemma}

\begin{proof}
We give 
a 1-interaction bounded QIP system $(P,V)$ that recognizes $Odd$.
Now, let $\Sigma=\{0,1\}$ and $\Gamma=\{\#,a\}$ be respectively the input alphabet and the communication alphabet for $(P,V)$. 
Let $Q=\{q_0,q_1,q_2,q_{acc},q_{rej,0},q_{rej,1}\}$ be the set of $V$'s inner states with $Q_{acc}=\{q_{acc}\}$ and $Q_{rej}=\{q_{rej,0},q_{rej,1}\}$. The protocol of the verifier $V$ 
is given as follows. With no query to a committed prover, $V$ continues to read the input symbols until the head scans $1$ in the input tape. 
When $V$ reads $1$, $V$ queries the symbol $a$ to a committed prover. 
If the prover returns $a$, then $V$ immediately rejects the input. Otherwise, the verifier checks whether 
the substring of the input after $1$ includes an odd number of $0$s. 
This check can be done by the verifier alone. Table \ref{fig:interaction} gives the formal description of $V$'s transitions. 
The honest prover $P$,  whenever receiving the symbol $a$ from the verifier, returns the symbol $\#$ and writes $a$ in the first blank cell of his private tape. Technically speaking, to make $P$ unitary, we need to map visible configuration $\qubit{\#}\qubit{y}$ for certain $y$'s not appeared yet in $P$'s private tape to superposition $\qubit{a}\qubit{\phi_{x,y}}$ with an appropriate vector $\qubit{\phi_{x,y}}$. By a right implementation, we can make $P$ a committed prover.

\begin{table}[ht]
\bs\begin{center}
\begin{tabular}{|lll|}\hline
$V_{\cent}\qubit{q_0}\qubit{\#}=\qubit{q_0}\qubit{\#}$ 
& $V_{0}\qubit{q_0}\qubit{\#}=\qubit{q_0}\qubit{\#}$ & 
$V_{1}\qubit{q_0}\qubit{\#}=\qubit{q_1}\qubit{a}$ \\  
$V_{\$}\qubit{q_0}\qubit{\#} =\qubit{q_{rej,0}}\qubit{\#}$ & 
$V_{0}\qubit{q_1}\qubit{\#}=\qubit{q_2}\qubit{\#}$ & 
$V_{1}\qubit{q_1}\qubit{\#}=\qubit{q_1}\qubit{\#}$ \\ 
$V_{\$}\qubit{q_1}\qubit{\#} =\qubit{q_{rej,1}}\qubit{\#}$ & 
$V_{0}\qubit{q_2}\qubit{\#} =\qubit{q_1}\qubit{\#}$ & 
$V_{1}\qubit{q_2}\qubit{\#} =\qubit{q_2}\qubit{\#}$ \\
$V_{\$}\qubit{q_2}\qubit{\#} =\qubit{q_{acc}}\qubit{\#}$ 
& $V_{0}\qubit{q_1}\qubit{a}=\qubit{q_{rej,0}}\qubit{\#}$ & 
$V_{1}\qubit{q_1}\qubit{a}=\qubit{q_{rej,0}}\qubit{\#}$ \\  \hline
\end{tabular}
\caption{Transitions of $V$ for $Odd$}\label{fig:interaction} 
\end{center}
\end{table}

We show that $(P,V)$ recognizes $Odd$ with probability $1$. Let $x$ be any input. 
First, consider the case where $x$ is in $Zero$. Assume that $x$ is of the form $0^m1y$, 
where $y$ contains an odd number of 0s. 
The honest prover $P$ erases $a$ that is sent from the verifier when $V$ reads $1$. Since $V$ can check whether $y$ includes an odd number of $0$s, $V$ accepts $x$ with certainty. 
Next, assume that $x\not\in Odd$. In the special case where $x\in\{0\}^*$, $V$ can reject $x$ with certainty with no query to a committed prover. 
Now, consider the remaining case where $x$ contains a $1$. Assume that $x$ is of the form $0^m1y$, where $y$ contains an even number of 0s. The verifier $V$ sends $a$ to a committed prover when he reads $1$. 
Note that $V$'s protocol is deterministic. To maximize the acceptance probability of $V$, a dishonest prover needs to return $\#$ to $V$ since, otherwise, $V$ immediately rejects $x$ in a deterministic fashion. 
Since $V$ can check whether $y$ includes an odd number of $0$s 
without making any query to the prover, for any committed prover $P^*$, $(P^*,V)$ rejects $x$ with certainty. 
Since the number of interactions in the protocol is at most $1$,  $Odd$ therefore belongs to $\qip_1^\#(1qfa)$, as requested. 
\end{proof}

As the second step, we prove Proposition \ref{zero-1qfa}. The language $Zero$ is known to be outside of $\mathrm{1QFA}$ \cite{KW97}; 
in other words, $Zero\not\in \qip^{\#}_{0}(1qfa)$. 
Proposition \ref{zero-1qfa} expands this result and shows that $Zero$ is not even in $\qip^{\#}_{1}(1qfa)$.

\begin{proposition}\label{zero-1qfa}
$Zero\not\in\qip^{\#}_{1}(1qfa)$.
\end{proposition}

Now, we begin with the proof of Proposition \ref{zero-1qfa}. Towards a contradiction, we first assume that a 1-iteration bounded QIP system $(P,V)$ with 1qfa verifier recognizes $Zero$ with error probability $\leq 1/2-\eta$ 
for a certain constant $\eta>0$. Let $Q$ and $\Gamma$ be respectively the set of $V$'s inner states and the communication alphabet. Write $\Sigma$ for our alphabet $\{0,1\}$ for simplicity. 
Without loss of generality, we assume henceforth that $V$ does not query at the time when it enters a halting inner state; in particular, the time when the head is scanning the endmarker $\$$. 

First, we introduce the notions of {\lq\lq}1-iteration condition{\rq\rq} and {\lq\lq}query weight.{\rq\rq} 
We fix an input $x$ and let $P'$ be any committed prover. For readability, we use the notation $Comp_{V}(P',x)$ to denote the computation of $(P',V)$ on the input $x$. Moreover, $PComp_{V}(P',x)$ denotes the partial computation obtained by executing the QIP protocol $(P',V)$ on any input whose prefix is $x$ while the head is reading $\cent x$ (\ie between the first step at $\cent$ 
and the step at which the head reads the rightmost symbol of $x$ and moves off $x$). When we consider a computation path, we understand that a computation path {\em terminates} either at a halting configuration or at a non-halting configuration $\xi$ of zero amplitude.

\begin{figure}[ht]
\begin{center}
\centerline{\psfig{figure=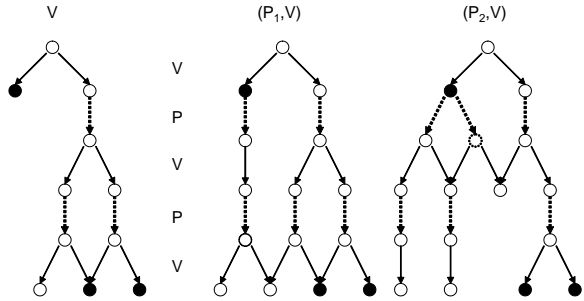,height=4.0cm}}
\caption{Example of a modified computation. The leftmost graph depicts the modified computation of $V$ on input $x$. The latter two graphs are computations of $V$ on $x$ using different provers $P_1$ and $P_2$. The black circles indicate query configurations whereas  the white circles indicate non-query configurations. The dotted circle is the place where prover $P_2$ forces $V$ to generate a new computation path 
that destructively interferes with an existing path in the modified computation of $V$.}\label{fig:query}
\end{center}
\end{figure}

For convenience, a committed prover $P'$ is said to satisfy {\em the 1-iteration condition} at $x$ with $V$ if, for any query configuration  $\xi$ of non-zero amplitude in $Comp_{V}(P',x)$, no other query configuration exists between $\xi$ and the initial configuration in the computation. Let $C^{(1)}_{x,V}$ be the collection of all committed provers $P'$ who satisfy the 1-iteration condition at $x$ with $V$. 
It is important to note that, whenever a prover in $C_{x,V}$ answers to $V$ with non-blank communication symbols 
with non-zero amplitude, $V$ must change these symbols back to blank immediately since, otherwise, $V$ is considered to make a second query. 
Choose any prover $P'$ in $C^{(1)}_{x,V}$ and consider the  computation $Comp_{V}(P',x)$. By introducing an extra projection, we modify $Comp_{V}(P',x)$ as follows. Whenever $V$ conducts a measurement, we then apply a projection, mapping onto the Hilbert space $\mathrm{span}\{\qubit{\#}\}$, to the communication cell. This projection makes all non-blank symbols collapse. If the communication cell is blank, then $V$ continues to the next step. 
Observe that this modified computation is independent of the choice of a committed prover. 
For this modified computation of $V$ on $x$, we use the notation $MComp_{V}(x)$. 
Figure \ref{fig:query} illustrates the difference between a modified computation and two computations with different provers. 
The {\em query weight $wt_{V}^{(x)}(y)$ of $V$ at $y$ conditional to $x$} is the sum of all the squared magnitudes of 
the amplitudes of query configurations, in $MComp_{V}(xy)$, where $V$ makes queries while reading $y$.  
For brevity, let $wt_{V}(y)=wt_{V}^{(\lambda)}(y)$, where $\lambda$ is the empty string. By its definition, a query weight ranges between $0$ and $1$ and satisfies that $wt_{V}(x) + wt^{(x)}_{V}(y) = wt_{V}(xy)$ for any $x,y\in\Sigma^*$.

Recall that $(P,V)$ is 1-iteration bounded and recognizes $Zero$. The following lemma holds for the query weight of $V$. In the lemma, {\em one round} in a computation comprises the following series of executions: a prover first applies his strategy including the return of the blank symbol (if not the first round) and $V$ then makes his move followed by a measurement. Note that the first round does not include a prover's move. In a modified computation, {\em one round} is similar but further includes an extra projection (described above) after $V$'s own measurement. 

\begin{lemma}\label{weight}
Let $P'$ be any committed prover and let $x,y$ be any strings.
\begin{enumerate}\vs{-2}
\item Any committed prover satisfies the 1-iteration condition at $x$ with $V$.
\vs{-2}
\item Any query configuration $\xi$ of non-zero amplitude at round $i$ in $Comp_{V}(P',x)$ must appear at the same round $i$ in $MComp_{V}(x)$ with the same amplitude for any $i\in[1,|x|+1]_{\integer}$. 
\vs{-2}
\item The query weight $wt_{V}^{(x)}(y)$ is greater than or equal to the sum of all the squared magnitudes 
of amplitudes of query configurations in $PComp_{V}(P',xy)$ while $V$'s head is reading $y$.  
\end{enumerate}
\end{lemma}

\begin{proof}
1) Take any committed prover $P'$. In the case where $x\not\in Zero$, since the QIP protocol $(P',V)$ makes at most 1 iteration on $x$,  $P'$ clearly satisfies the 1-iteration condition at $x$ with $V$. In contrast, assuming that $x\in Zero$, consider the partial computation $PComp_{V}(P',x)$. Note that this partial computation is also a partial computation of $(P',V)$ on $x1$. Since $x1\not\in Zero$, $P'$ must satisfy the 1-iteration condition at $x1$ with $V$. Therefore, $P'$ satisfies the 1-iteration condition also at $x$.   

2) We prove the claim by induction on $i\in[1,|x|+1]_{\integer}$. The basis case $i=1$ is trivially true since there is no prover's strategy. Consider the $i$th round. Let $\xi$ be any query configuration of non-zero amplitude occurring at round $i$ in $Comp_{V}(P',x)$. Note from the first claim that every committed prover satisfies the 1-iteration condition at $x$ with $V$.

We first show that $\xi$ appears with non-zero amplitude at round $i$ in $MComp_{V}(x)$. Assume otherwise that $\xi$ appears at round $i$ in $Comp_{V}(P',x)$ but not in $MComp_{V}(x)$. 
This implies that, at a certain early round, as a response to a query configuration $\eta$ in $Comp_{V}(P',x)$ of non-zero amplitude, $P'$ forces $V$ to generate $\xi$ with non-zero amplitude later at the round $i$ since, otherwise, $V$ generates $\xi$ with no query and $\xi$ is therefore in $MComp_{V}(x)$, a contradiction. Since $\eta$ and $\xi$ are in the same computation path, this clearly violates the 1-iteration condition of $P'$. As a consequence, $\xi$ must appear with non-zero amplitude at round $i$ in $MComp_{V}(x)$. 

Next, we show that $\xi$'s amplitude in $Comp_V(P',x)$ is the same as in $MComp_{V}(x)$. Towards a contradiction, we assume that the amplitudes of $\xi$ in $Comp_{V}(P',x)$ and in $MComp_{V}(x)$ are different. Now, consider all computation paths that reach $\xi$. Note that, if such a path contains no query configuration (other than $\xi$), this path must appear in $MComp_{V}(x)$. There are two cases to discuss: either a new computation path leading to $\xi$ is added or an existing computation path to $\xi$ is annihilated. 

{\sf (Case 1)} Consider any computation path $\gamma$ leading to $\xi$ in $Comp_{V}(P',x)$ 
whose amplitude contributes to the difference of $\xi$'s amplitudes in $Comp_{V}(P',x)$ and in $MComp_{V}(x)$. 
Such a path $\gamma$ should not be present in $MComp_{V}(x)$. 
The 1-iteration condition of $P'$ implies that, since $\xi$'s  amplitude is not $0$, 
the path $\gamma$ cannot contain any query configuration of non-zero amplitude before reaching $\xi$.  Hence, the path $\gamma$ must be in $MComp_{V}(x)$, a contradiction.

{\sf (Case 2)} The remaining case is that, at an early round, $P'$ forces $V$ to generate a number of computation paths 
that destructively interfere with an existing computation path $\delta$ leading to $\xi$ in $MComp_{V}(x)$. This interference annihilates the path $\delta$, which causes the change of $\xi$'s amplitude in $Comp_{V}(P',x)$. 
Figure \ref{fig:query} illustrates this case. 
We modify the strategy of $P'$ by changing its amplitudes (but not the tape/communication symbols) 
so that $\delta$ narrowly survives. Note that such a modification is possible because $V$ moves exactly 
in the same way as before and therefore the modification does not incur any change of the computation $Comp_{V}(P',x)$ 
except for the amplitude distribution. 
As a result, the path $\delta$ connects two query configurations of non-zero amplitudes. 
This contradicts the first claim; namely, the 1-iteration condition of any committed prover. 

In either case, we reach a contradiction. Therefore, the claim holds.

3) This follows directly from the second claim. 
\end{proof}

We continue the proof of Proposition \ref{zero-1qfa}. Now, consider the value $\nu$ defined as the supremum, over all strings $w$ in $Zero$, of the query weight of $V$ at $w$. Observe that $0\leq \nu\leq 1$ by Lemma \ref{weight}.
We examine the two cases $\nu=0$ and $\nu>0$ separately. For readability, we omit the letter $V$ whenever it is clear from the context.

{\sf (Case 1: $\nu=0$)}\hs{2} 
Obviously, $wt(w)=0$ for all $w\in Zero$. 
Toward a contradiction, it suffices to give a bounded-error 1qfa that recognizes $Zero$ since $Zero\not\in \mathrm{1QFA}$. Let $P_I$ be the committed prover who applies only the identity operator at every step. 
The desired 1qfa $M$ behaves as follows. 
On input $x$, $M$ simulates $V$ on $x$ with the {\lq\lq}imaginary{\rq\rq} prover $P_I$ by maintaining the content of the communication cell 
as an integrated part of $M$'s inner states.  
This is possible by defining $M$'s inner state $(q,\sigma)$ to reflect both $V$'s inner state $q$ 
and a symbol $\sigma$ in the communication cell. 
Now, we claim that $M$ recognizes $Zero$ with bounded error. 
If input $x$ is in $Zero$, then, since any communication with a prover has the zero amplitude, 
$M$ correctly accepts $x$ with probability $\geq 1/2+\eta$. 
Similarly, we can verify that, if $x$ is not in $Zero$, $M$ rejects $x$ with probability $\geq 1/2+\eta$  because $(P_{I},V)$ must reject $x$ with the same probability.
Therefore, $M$ recognizes $Zero$ with error probability $\leq 1/2-\eta$, as requested.  

{\sf (Case 2: $\nu>0$)}\hs{2} 
Recall that the notation $P_{w}$ refers to the strategy of $P$ on input $w$. Note that, for every real number $\gamma\in(0,\nu]$, there exists a string $w$ in $Zero$ such that $wt(w) \geq \nu-\gamma$. 
For each $y\in\Sigma^*$, set $\gamma_y=\min\{\eta^2/16(|y|+1)^2,\nu\}$ 
and choose the lexicographically minimal string $w_{y}\in Zero$ such that $wt(w_y) \geq \nu - \gamma_{y}$.

For each $y\in\Sigma^*$, define the new prover $P'_{y}$ that behaves on input $w_{y}y01^m$ for every $m\in\nat^{+}$ 
in the following fashion: 
$P'_{y}$ takes the strategy $P_{w_{y}y0}$ 
while $V$'s head is reading $\cent w_{y}$ and then $P'_{y}$ behaves as $P_{I}$ (\ie applies the identity operator) while $V$ is reading the remaining portion $y01^m\$$. For readability, we abbreviate $w_{y}y$ as $\tilde{y}$. We then claim the following. 

\bs
\n{\bf Claim.}\hs{1}  
{\em For any string $y\in\Sigma^*$, $p_{acc}(\tilde{y}0,P'_{y}) \geq 1/2+\eta/2$.} 
\vs{-1}

\begin{proofof}{Claim}
Let $y$ be an arbitrary input string. 
Note that the protocol $(P'_{y},V)$ works in the same way as $(P_{\tilde{y}0},V)$ 
while $V$ is reading $\cent w_{y}$. 
Consider $wt^{(w_y)}(y0)$. Note that $wt(w_y) + wt^{(w_y)}(y0)\leq \nu$. 
It thus follows that $wt^{(w_y)}(y0)\leq \gamma_{y}$ using the inequality that $wt(w_y)\geq \nu - \gamma_{y}$. 
Lemma \ref{weight}(3) implies that, for any committed prover $P^*$, 
$wt^{(w_y)}(y0)$ bounds the sum of all the squared magnitudes of query configurations, 
while the head is reading $y0$, in the computation of $(P^*,V)$ on the input $w_{y}y0$. 
Therefore, a simple calculation (as in, \eg \cite[Lemma 9]{Yam03}) shows that
\[ 
|p_{acc}(\tilde{y}0,P'_{y}) - p_{acc}(\tilde{y}0,P_{\tilde{y}0})|
\leq 2\left(wt^{(w_y)}(y0)\right)^{1/2}|y0| \leq 2\sqrt{\gamma_{y}}(|y|+1) \leq \eta/2. 
\]
Since $p_{acc}(\tilde{y}0,P_{\tilde{y}0})\geq 1/2+\eta$, 
it follows that $p_{acc}(\tilde{y}0,P'_{y}) 
\geq (1/2+\eta)- \eta/2 \geq 1/2+\eta/2$. 
\end{proofof}

Recall the set $Q$ of inner states and the communication alphabet $\Gamma$. Set $d=|Q||\Gamma|$ for brevity. Using Lemma \ref{one-qfa-bound}, for each $y\in\Sigma^*$, there exists a $(|\tilde{y}0|+2,d)$-bounded QIP system $(P^{(1)}_{y},V)$ that simulates $(P'_{y},V)$ on input $\tilde{y}0$. 
The initial superposition is $\qubit{q_0,\#,\#^{d}}$, where we omit the qubits representing the head position of $V$ because $V$ is a 1qfa verifier. Let $\VV=\mathrm{span}\{\qubit{q}\mid q\in Q\}$, let $\MM=\mathrm{span}\{\qubit{\sigma}\mid \sigma\in \Gamma\}$, and let $\PP$ be the $d$-dimensional Hilbert space representing the prover's private tape.
Let $\qubit{\psi_y}$ be the superposition in the global configuration space $\VV\otimes\MM\otimes\PP$ obtained 
just after $V$'s head moves off the right end of $\cent \tilde{y}0$ and $P^{(1)}_y$ replies to $V$. 
For each number $n\in\nat^{+}$, consider a $(|\tilde{y}0|+2+n,d)$-bounded QIP system $(P^{(2)}_{y,n},V)$ where 
$P^{(2)}_{y,n}$ simulates $P'_{y}$ while reading $\cent \tilde{y}0$ and applies the identity operator while reading $1^n\$$. 
Noting that the prover $P'_{y}$ does nothing after $V$ have read $\cent w_{y}$, 
we can verify that $(P^{(2)}_{y,n},V)$ simulates $(P'_{y},V)$ on the input $\tilde{y}01^n$. 
Letting $\mu=\mathrm{inf}_{y\in\Sigma^*}\{\|\qubit{\psi_y}\|\}$, we consider the two subcases $\mu\leq \eta/4$ and $\mu>\eta/4$.

{\sf (Subcase a: $\mu\leq \eta/4$)}\hs{2} 
There exists a string $y$ such that $\mu\leq \|\qubit{\psi_y}\|<\mu+\eta/4\leq \eta/2$. 
This means that, after reading $\cent \tilde{y}0$, the halting probability of $V$ increases by no more than $(\eta/2)^2$. 
Consider the input $\tilde{y}01$. Since $p_{acc}(\tilde{y}0,P^{(1)}_y)\geq 1/2+\eta/2$, 
it follows that $p_{acc}(\tilde{y}01,P^{(2)}_{y,1})\geq (1/2+\eta/2)-\|\qubit{\psi_{y}}\|^2 \geq 1/2$. 
However, this contradicts our assumption that, for any committed prover $P^*$, $(P^*,V)$ accepts $\tilde{y}01$ with probability $\leq 1/2-\eta<1/2$. 

{\sf (Subcase b: $\mu> \eta/4$)}\hs{2} 
Let $\epsilon$ be any sufficiently small positive real number and choose a string $y$ 
such that $\|\qubit{\psi_y}\|\in[\mu,\mu+\epsilon)$. 
The superposition of global configurations obtained after the operation of the protocol $(P^{(2)}_{y},V)$ on $\tilde{y}01^j$ 
just before $V$ scans $\$$ becomes $(P_{I}E_{non}V_1)^j\qubit{\psi_y}$, 
where $V_b$ ($b\in\check{\Sigma}$) is the unitary operation of $V$ when $V$ is scanning the symbol $b$ and $P_{I}$ is the identity operation of a prover. 
For convenience, write $W$ for $P_I E_{non}V_1$. 
For any integer $j\geq 1$, $\mu\leq \|W^j \qubit{\psi_y}\| <\mu+\epsilon$. 
By a similar analysis in \cite{KW97} (see also \cite[Lemma 4.1.12]{Gru99}), 
there exist a constant $c>0$ independent of $\epsilon$ and a number $m\in\nat^{+}$ such that  
$\|\qubit{\psi_y}-W^m\qubit{\psi_y}\|<c\cdot \epsilon^{1/4}$. 
{}From this inequality follows 
\[
|p_{acc}(\tilde{y}0,P^{(1)}_{y}) - p_{acc}(\tilde{y}01^m,P^{(2)}_{y,n})|
\leq \|V_{\$}\qubit{\psi_y} - V_{\$}W^m\qubit{\psi_y}\| = \|\qubit{\psi_y} - W^m\qubit{\psi_y}\| < c\epsilon^{1/4}, 
\]
where the first inequality is obtained as in the proof of Lemma \ref{mo-weak}. We thus obtain the upper bound that $|p_{acc}(\tilde{y}0,P'_{y}) - p_{acc}(\tilde{y}01^m,P'_{y})|\leq c\epsilon^{1/4}$. 
Let $\epsilon=\left(\frac{\eta}{2c}\right)^4$. 
Since $p_{acc}(\tilde{y}0,P'_{y})\geq 1/2+\eta/2$, 
it follows that $p_{acc}(\tilde{y}01^m,P'_{y})\geq (1/2+\eta/2)-c\epsilon^{1/4} = 1/2$.
This contradicts our assumption that $(P^*,V)$ accepts $\tilde{y}01^m$ with probability $\leq 1/2-\eta/2 < 1/2$ for any committed prover $P^*$. 
Therefore, $Zero\not\in\qip^{\#}_{1}(1qfa)$, as requested.  This completes the proof of Proposition \ref{zero-1qfa}.

Since a 1qfa verifier cannot remember the number of queries,  
we may not directly generalize the proof of Theorem \ref{query-once} to claim that $\qip^{\#}_k(1qfa)\neq \qip^{\#}_{k+1}(1qfa)$ for any constant $k$ in $\nat^{+}$. 
Nevertheless, we still conjecture that this claim holds. 

\section{Future Directions}

There have been a surge of interests in QIP systems 
\cite{CHTW04,KW00,KM03,Wat03,Yao03} partly 
because a QIP system embodies an essence of quantum 
computation and communication. 
Our research on weak-verifier QIP systems was inspired by 
the work of Dwork and Stockmeyer \cite{DS92}, 
who extensively studied $\ip(2pfa)$ and $\am(2pfa)$. Having 
started with our basic qfa-verifier QIP systems, 
we have discussed several variants of restricted QIP systems 
and have demonstrated strengths and weaknesses 
of these QIP systems. Nonetheless, the theory of weak-verifier 
QIP systems is still vastly uncultivated. 
The development of new proof techniques is needed to settle 
down, for instance, all the pending questions left in this 
paper. 
We strongly hope that further research will unearth the crucial 
characteristics of the QIP systems. 

This final section discusses six important directions that 
lead to fruitful future research on weak-verifier QIP systems. 
\begin{itemize}
\item{\bf Modifying Verifier's Ability.}
Our verifier is a quantum finite automaton against a mighty 
prover who can apply any unitary operation. 
This paper has dealt with only three major qfa's: mo-1qfa's, 
1qfa's, and 2qfa's. It is important to study the nature of 
quantum interactions between provers and different types of verifiers. 
There have been several variants of 2qfa's proposed in the literature. 
For instance, Amano and Iwama \cite{AI99} studied so-called a 1.5qfa, 
which is a 2qfa whose head never moves to the left. Recently, 
Ambainis and Watrous \cite{AW02} considered a 2qfa whose head 
move is particularly classical. Instead of restricting the 
ability of qfa's, we can supplement an additional device to 
gain more computational power of qfa's. 
As an example, Golovkins \cite{Glov} lately studied a qfa that 
is equipped with a pushdown stack. Using these qfa models as 
verifiers, we need to conduct a comprehensive study on the 
corresponding QIP systems.
\vs{-2}
\item{\bf Curtailing Prover's Strategy.}
Another direction is to limit the prover's power. Instead of 
strengthening a verifier, for instance, we can restrict the 
size of the prover's strategy. Having already seen in Lemma 
\ref{prop2-KM03}, without diminishing the recognition power, 
we can limit the size of prover's private tape space to the 
size of the verifier's visible configuration space. If we 
further constrain the prover's strategy, how powerful is the 
corresponding QIP system? 
In the 1990s, Condon and Ladner \cite{CL92} studied IP systems 
with restricted provers who take only polynomial-size strategy. 
They showed that, with polynomial-size strategy, 
the IP systems with polynomial-time PTM verifiers exactly 
characterize Babai's class $\mathrm{MA}$. Analogously, for instance, 
we can consider the QIP systems in which 2qfa verifiers play 
against $O(\log\log{n})$-space bounded provers. 
Such QIP systems still recognize certain non-regular languages.
\vs{-2}
\item{\bf Communicating through a Classical Channel.}
We may understand our QIP protocol as a 2-party communication 
protocol exchanging messages through a quantum channel. 
Recall that a classical prover performs only a unitary operation 
of entries either 0 or 1. 
Seen as a communication protocol, we instead restrict a 
communication channel between two players, a prover and a 
verifier, to be classical. 
Such a communication may be realized by performing a measurement 
on  the communication cell just before each player makes an access 
to the cell. The communication cell then becomes a probabilistic 
mixture of classical states. It is, nonetheless, unclear whether 
this QIP system is as powerful as our classical-prover QIP system.
\vs{-2}
\item{\bf Using Prior Entanglement.}
Quantum entanglement is of significant importance in quantum computation 
and communication. The EPR pair\footnote{The EPR pair is the 2-qubit 
quantum state $\qubit{\Phi^+}=(\qubit{00}+\qubit{11})/\sqrt{2}$, 
which was proposed by Einstein, Podolsky, and Rosen in 1935.}, for 
instance, is used to teleport a quantum state using a quantum correlation 
between two qubits. 
Consider the case where a verifier shares limited prior 
entanglement with a prover in such a way that, 
before the start of a QIP protocol, a certain number of the verifier's 
inner states and a finite segment of the prover's private tape are 
entangled in a predetermined manner. 
This simple model of limited prior entanglement, nevertheless, 
does not enrich the computational resource of the QIP systems 
because, similar to the proof of $\qip(1qfa)=\mathrm{REG}$, we 
can prove that the aforementioned limited prior entanglement 
makes the corresponding QIP systems recognize only regular languages. 
Therefore, other types of models are needed to explore the 
usefulness of prior entanglement.
\vs{-2}
\item{\bf Playing against Multiple Provers.}
A natural extension of our basic QIP systems is obtained by providing each QIP system with multiple provers against a single verifier. In the polynomial-time setting, Kobayashi and Matsumoto \cite{KM03} studied the QIP systems in which a uniform polynomial-size quantum-circuit verifier plays against multiple provers. These provers may further share prior entanglement among them (but not with a verifier). 
Multiple-prover QIP systems of Kobayashi and Matsumoto are shown to characterize the complexity class $\mathrm{NEXP}$ \cite{KM03}. 
In a classical case, Feige and Shamir \cite{FS92} constructed a 2-prover IP system with a 2pfa verifier (using the model of Dwork and Stockmeyer) for each recursive language. Naturally, we expect the multiple-prover QIP systems with qfa verifiers to demonstrate a similar increase in power over the single-prover QIP systems.
\vs{-2}
\item{\bf Making Knowledge-Based Interactions.} 
Lately, a great attention has been paid to a quantum zero-knowledge proof systems 
(QZKP systems, in short) \cite{Kob03,Wat02}. 
As a followup to their 2pfa-verifier IP systems, 
Dwork and Stockmeyer also studied zero-knowledge proof systems played between provers and 2pfa verifiers \cite{DS92-2}. 
It is desirable to develop a theory of QZKP systems with qfa verifiers in connection to quantum cryptography.
\end{itemize}

\paragraph{Acknowledgment.} The first author is 
grateful to Hirotada Kobayashi for a detailed presentation of his result.

\bibliographystyle{alpha}

\end{document}